\documentclass[11pt]{article}

\usepackage[T1]{fontenc}
\usepackage[utf8]{inputenc}

\usepackage{color}
\usepackage{latexsym}
\usepackage{amsmath,amsfonts,amssymb,theorem,euscript,array,enumerate,mathrsfs,cases,bbm}
\usepackage{graphicx,subfigure}
\usepackage{dsfont}
\usepackage{geometry}

\usepackage{epsfig,epstopdf}
\usepackage{hyperref}
\hypersetup{
    colorlinks,
    citecolor=blue,
    filecolor=blue,
    linkcolor=blue,
    urlcolor=blue
}
\usepackage[normalem]{ulem}
\usepackage{cancel}

\usepackage{float,caption}

\DeclareGraphicsExtensions{.png,.jpg,.bmp,.eps,.pdf}

\newtheorem{Theorem}{Theorem}[section]

\newtheorem{Proposition}[Theorem]{Proposition}

\newtheorem{Lemma}[Theorem]{Lemma}
\newtheorem{Corollary}[Theorem]{Corollary}
\newtheorem{Remark}[Theorem]{Remark}

\numberwithin{equation}{section}

\def\esssup_#1{\underset{#1}{\mathrm{ess\,sup\, }}}
\def\essinf_#1{\underset{#1}{\mathrm{ess\,inf\, }}}
\def\argmax_#1{\underset{#1}{\mathrm{arg\,max\, }}}
\def\argmin_#1{\underset{#1}{\mathrm{arg\,min\, }}}

\def \trans{^{\scriptscriptstyle{\intercal}}}

\def \N{\mathbb{N}}
\def \R{\mathbb{R}}

\def \E{\mathbb{E}}
\def \F{\mathbb{F}}

\def \P{\mathbb{P}}

\def \Ac{{\cal A}}
\def \Bc{{\cal B}}

\def \Fc{{\cal F}}

\def \Pc{{\cal P}}

\def \ep{\hbox{ }\hfill$\Box$}

\def\beqs{\begin{eqnarray*}}
\def\enqs{\end{eqnarray*}}
\def\beq{\begin{eqnarray}}
\def\enq{\end{eqnarray}}

\allowdisplaybreaks

\newcolumntype{M}[1]{>{\centering\arraybackslash}m{#1}}
\newcolumntype{P}[1]{>{\centering\arraybackslash}p{#1}}

\graphicspath{{./fig/}}

\begin{document}

\title{Equilibrium price in intraday electricity markets\footnote{This study was supported by FiME (Finance for Energy Market Research Centre) and
the ``Finance et D\'eveloppement Durable - Approches Quantitatives'' EDF - CACIB Chair.}  
}

\author{
Ren\'e AID\footnote{LEDa, Universit\'e Paris Dauphine, PSL research university, \sf rene.aid at dauphine.fr}
\qquad
Andrea COSSO\footnote{Department of Mathematics, University of Bologna, \sf andrea.cosso at unibo.it}
\qquad
Huy\^en PHAM\footnote{LPSM, Universit\'e de Paris, \sf pham at lpsm.paris}
}

\maketitle

\date{}

\begin{abstract}
We formulate an equilibrium model of intraday trading in electricity markets. Agents face balancing constraints between their customers consumption plus intraday sales and their production plus intraday purchases. They have continuously updated forecast of their customers consumption at maturity with decreasing volatility error. Forecasts are prone to idiosyncratic noise as well as common noise (weather). Agents production capacities are subject to independent random outages, which are each modelled by a  Markov chain. 
The equilibrium price is defined as the price that minimises trading cost plus imbalance cost of each agent and satisfies the usual market clearing condition.  Existence and uniqueness of the equilibrium are proved, and we 
show that the equilibrium price and the optimal trading strategies are martingales. The main economic insights are the following.  (i) When there is no uncertainty on generation, it is shown that the market price is a convex combination of forecasted marginal cost of each agent, with deterministic weights. 
Furhermore,  the equilibrium market price follows Almgren and Chriss's model and we identify the fundamental part as well as the permanent market impact. It turns out that heterogeneity across agents is a necessary condition for the Samuelson's effect to hold. (ii) When there is production uncertainty, the price volatility becomes stochastic but converges to the case without production uncertainty when the number of agents increases to infinity. Further, on a two-agent case, we show that the potential outages of a low marginal cost producer reduces her sales position.
\end{abstract}

\vspace{5mm}


\vspace{5mm}


\clearpage


\section{Introduction}

Because electricity cannot be stored, the development of competitive electricity markets has led to the introduction of intraday markets. The purpose of these markets whose time-horizon does not exceed more than 36 hours is to allow electricity market players to balance their position between their customers consumption and their generation for each hour of the day and avoid expensive imbalance costs. The development of intermitent renewable electricity generation have increased the interest of both market players and academics for these markets. Indeed, because of the high uncertainty of wind generation production, the need of short-term balancing mechanism has increased. Empirical studies of intraday market show an increase in trading volume in the last years and convergent stylised facts about liquidity and volatility of intraday prices; liquidity as measured by market depth and volatility both increase with time closer to delivery (see Kiesel et. al. (2017) \cite{Kiesel17}, Balardy (2018) \cite{Balardy18}, Kremer et. al. \cite{Kremer20}, Glas et. al. (2020) \cite{Glas20}).

This phenomenom is known as the {\em Samuelson's effect}, since it was first posited and explained by Samuelson (1965) \cite{Samuelson65}. This effect states that the volatility of futures prices contract increases as time gets closer to delivery.  And indeed, trading in intraday market consists in trading during less than 36 hours a futures contract for delivery for a maturity given by a fixed hour. This pattern of increasing volatility of electricity futures prices has been found for other maturities. Jaeck and Lautier (2016)  \cite{Jaeck16} finds that Samuelson's effect holds on all tested electricity markets (German, Nordic, US, Autralia) for monthly contract delivery. 

In Samuelson's paper, the effect is obtained as the consequence of two hypothesis: the spot price is mean-reverting and the futures price is the conditional expectation of the spot price. Further, for Bessembinder et. al. (1996) \cite{Bessembinder96}, mean-reversion in the context of commodity prices is linked to storability: when prices are high, agents reduce storage making the price decrease and vice versa, resulting in a mean-reversion effect. In the context of electricity, the storage hypothesis does not seem appropriate to explain the effect. Further, all prices of storable commodities do not exhibit this behaviour (see Jaeck and Lautier (2016) \cite{Jaeck16}). Anderson and Danthine (1983) \cite{Anderson83} and Anderson (1985) \cite{Anderson85} formulated a more general hypothesis, named {\em the state variable hypothesis} to explain why some commodity prices exhibith the Samuelson's effect and others do not. They state that the monotonocity (if any) of the volatility of futures prices depends on the way uncertainty on the equilibrium between demand and supply is resolved. In particular, in the case where volatility of demand uncertainty decreases with time, the futures price volatility may decrease. This is the case of intraday electricity market (demand forecasts error tend to decrease quickly with delivery) and nevertheless, intraday prices still exhibit an increase of volatility. Another explanation for the occurence or not of the Samuelson's effect has been formulated by Hong (2000) \cite{Hong00}. Hong (2000) shows in an equilibrium model of commodity futures trading how information asymmetry may also sustain violation of Samuelson's hypothesis. In the case of electricity markets and in particular intraday markets, both the energy regulator and the European financial regulation under REMIT compels producers and retailers to provide immediate communication to the market operator of any information that may affect the equilibrium between supply and demand {\em before} taking appropriate trading positions. These regulations are intented to reduce as much as possible asymmetry of information between players and yet, the intraday electricity prices still exhibit a pattern of increasing volatility. More recently Féron et. al. (2020) \cite{Feron20} models an intraday market Nash equilibrium in the context of identical agents trading at a fundamental price plus a liquidity premium while the market price is defined {\em à la} Almgren and Chriss by a fundamental price plus a linear permanent impact induced by the average inventory level (because there is no market clearing condition, the average inventory is non-zero). In this context, they recover the Samuelson's effect as the result of strategic behaviour of agents.

In this paper, we develop an equilibrium model of intraday trading for a fixed hour of delivery $T$ with the purpose to explain and understand the optimal trading strategies of market players as well as the market price dynamics. We consider that the market is composed of $N$ agents $i$ having each a forecast $D^i_t$ at time $t \in [0,T]$ of their customers consumption at time $T$. We suppose that the volatility $\sigma^i_t$ of the demand forecast is deterministic and  
decreases with time, capturing the empirical evidence of increasing demand forecast with time to delivery (see Nedellec et. al. \cite{Nedellec14}). The demand forecast off each agent is affected both by an individual brownian noise and a collective brownian noise with correlation $\rho_i$, reflecting the dependence of market players to weather and global economic conditions. Further, each agent is endowed with a generation capacity with linear marginal cost of coefficient $\beta^i_t$. Further, $\beta^i_t$ evolves following a Markov chain, capturing by this way the possibility of power plant outages driving the marginal cost of an agent from a low to a high cost. Agents can buy or sell power for delivery at time $T$ at the market price plus a liquidity premium  $\gamma_i q^i_t$  proportional to the trade $q^i_t$. This premium translates the potential diffferent market access cost of agents. The objective of each agent is to minimise the expected total trading cost plus the costs of imbalance $\eta_i (D^i_T - X^i_T - \xi^i_T)^2$ where $\eta_i$ represents agent's $i$ own perception of the cost of imbalance, $X^i_T$ and $\xi^i_T$ are the inventory and the production of the agent at $T$. Although  the cost of imbalance is fixed by the electricity network operator and is the same for any market player, we allow for different evaluation of the cost of imbalance, translating the possibility that some players may have strong reluctance for imbalances while other might not care as much. All information is considered public. A market equilibrium is defined as trading strategies and a market price such that each agent has minimised her criteria and the market clears for the market price. This model owes agent's features to Aïd et. al. (2016) \cite{Aid16} and Tan and Tankov (2018) \cite{Tan18}.

In this framework, we obtain the following results. We prove existence and uniqueness of the equilibrium. The proof is based on the martingale optimality principle in stochastic control, and existence of solution to  backward stochastic differential equations (backward SDEs) with jumps, for which we provide a complementary existence result to Becherer (2002) \cite{Becherer02}.  We show that both the equilibrium price and the optimal trading strategies are martingales, and they are characterized in terms of a coupled system of  forward-backward SDE that we solve with explicit formulae. 

In the case where there is no production cost uncertainty, we observe  that the market price is a convex combination of the forecasted marginal cost of each agent where the weights are deterministic functions of time. The optimal trading rate of each agent consists in comparing her forecasted marginal cost to the market price and to take position accordingly, i.e. to sell (resp. to buy) if it is lower (resp. higher). Although simple, this strategy is commonly used in intraday electricity trading desks of power utilities. Further, we show that the equilibrium price has the form of Almgren and Chriss model \cite{Almgren01}. We identify the fundamental part of the price as the average forecasted marginal cost to satisfy the demands and identify the market permanent impact of each agents. Permanent market impacts are deterministic function of time with a monotony depending on the agent. If all agents are identical, the market equilibrium reduces to its fundamental component because of the market clearing condition.  The closed-form expression derived  for the price and the trading strategies allows us  to provide insight on the dynamics of the price volatility defined as the quadratic variation of the price. If all agents are identical, the price volatility monotonicity is fully determined by the volatility of the demand forecasts. In our case where the demand forecasts volatilities are decreasing in time, it implies that if the Samuelson's effect is to hold, agents must be heterogeneous. Thus, the Anderson and Danthine state variable hypothesis is not sufficient to explain increasing price volatility in a context of decreasing demand forecast error. We provide numerical illustrations where the mixing of agents of two different types allows to have decreasing or increasing volatility functions depending on the proportion of the agent's type. Further, heterogeneity of agents as expressed by their marginal cost, market access quality and dependence to weather can be observed. Thus,  explaining price volatility by heterogeneity leads to testable predictions.

In the case where there is production cost uncertainty, we show that if the number of agents is large, the equilibrium tends to the case of no production cost uncertainty because of the independence of jumps in the Markov chains between agents. Further, in the case of two players where the second player is affected by a potential jump that will switch her production cost from a lower marginal cost to a higher marginal cost compared to the first player, we observe that she moves from a selling position to a buying position as the probability of jumps increases. Although limited, this result gives credit to idea of precautionary position when entering in intraday market, i.e. selling less than the total quantity of marginal cost lower than the price.

The paper is structured as follows. Section~\ref{S:Model} describes precisely the model. Section~\ref{S:Martingale} provides the main results in terms of optimal strategies of each player for a given price process. Section~\ref{S:Equilibrium} provides the market equilibrium characterization by solving explicitly the coupled system of forward-backward SDE,  and the martingale properties of the equilibrium price. Section~\ref{S:NoJump} gives the description of the market equilibrium in the case of no production uncertainty while section~\ref{S:Jumps} provides the result in the other case. 


\section{The equilibrium model}
\label{S:Model}

We consider an economy with $N\in\N\backslash\{0\}$ power producers which can buy/sell energy on an intraday electricity market. Their purpose is to satisfy the demand of their customers at a given fixed time $T$, minimizing trading costs.

\subsection{Single agent optimal execution problem}

Following \cite{Aid16}, we formulate the optimization problem of a single agent $i\in\{1,2,\ldots,N\}$ in the economy on a finite time horizon $T>0$. We begin introducing some notations.

Consider a complete probability space $(\Omega,\Fc,\P)$ and a finite set $E\subset(0,+\infty)$ of cardinality $M$, where $M$ is a positive integer. We fix the following quantities at the initial time $t=0$:
\begin{itemize}
\item the initial demand forecasts of the agents $(d^1_0,d^2_0,\ldots,d^N_0)\in\R^N$;
\item the initial production capacities $(e^1_0,e^2_0,\ldots,e^N_0)\in E^N$;
\item the initial net positions of the agents of sales/purchases of electricity in the intraday electricity market $(x^1_0,x^2_0,\ldots,x^N_0)\in\R^N$.
\end{itemize}
On $(\Omega,\Fc,\P)$ we consider $N+1$ independent real-valued Brownian motions $(W_t^0)_{t\geq0}$,$(W_t^1)_{t\geq0}$,$\,\ldots\,$, $(W_t^N)_{t\geq0}$ and $N$ independent continuous-time homogenous Markov chains $(\beta_t^1)_{t\geq0}$,$\,\ldots\,$, $(\beta_t^N)_{t\geq0}$. We assume that $(W^0,W^1,\ldots,W^N)$ and $(\beta^1,\ldots,\beta^N)$ are independent. Moreover, every Markov chain 
$\beta^i$ is supposed to have finite state space $E$, starting point $e^i_0$ at time $t=0$: it represents the uncertainty over time on the production capacity of agent $i$.   We denote by $\Lambda_i=(\lambda_i(e,e')\colon e,e'\in E)$ the intensity matrix of $\beta^i$. We also denote by $\F=(\Fc_t)_{t\geq0}$ the augmentation of the filtration generated by $(W^0,W^1,\ldots,W^N)$ and $(\beta^1,\ldots,\beta^N)$. Finally, $\Pc$ denotes the predictable $\sigma$-algebra on $\Omega\times[0,T]$ associated with $\F$.

The demand forecast $D^i$ of agent $i$ evolves on $[0,T]$ according to the equation
\begin{equation}\label{Di}
D_t^i \ = \ d_0^i + \mu_i\,t + \int_0^t\sigma_s^i\,\Big(\rho_i\,dW_s^0 + \sqrt{1 - \rho_i^2}\,dW_s^i\Big),
\end{equation}
where $\mu_i\in\R$, $\rho_i\in[-1,1]$ and $\sigma^i\colon[0,T]\rightarrow\R$ is a decreasing function of time. In applications, we use the following form for $\sigma^i_t$
\begin{equation}\label{sigma_i}
\sigma_t^i \ = \ \sqrt{\sigma_i^2\,(T-t) + \sigma_0^2}, \qquad 0\leq t\leq T,
\end{equation}
for some $\sigma_i>0$ and $\sigma_0>0$. The volatility $\sigma^i_t$ captures the main features of demand forecast error at the intraday time horizon: it decreases at the root of the time to maturity and may have a residual term $\sigma_0$, capturing the incompressible demand forecast errors (see for instance \cite{Nedellec14} and the references within for an overview of the field of short term electricity demand forecast). Further, the dynamics of $D^i$ takes into account the potential common dependence of realised demands to weather conditions. In order to satisfy the terminal demand $D_T^i$, agent $i$ have the two following possibilities.

\begin{itemize}
\item\emph{Power production.} The agent can choose to product a quantity $\xi_i$, facing at the terminal time $T$ the cost
\begin{equation}\label{cost}
c_i(\xi_i) \ = \ \frac{1}{2}\,\beta_T^i\,\xi_i^2.
\end{equation}
\item\emph{Trading in intraday electricity market.} Let $X_t^{i,q^i}$ denote the agent net position of sales/purchases of electricity at time $t\in[0,T]$, delivered at the terminal date $T$, which is given by
\[
X_t^{i,q^i} \ = \ x_0^i + \int_0^t q_s^i\,ds,
\]
where $q^i$, called the trading rate, is chosen by the agent.
\end{itemize}
We define an admissible pair of controls for each agent $i$ as a pair $(q,\xi)$ in $\Ac^q\times\Ac^{\xi,+}$, where
\begin{align*}
\Ac^q \ &= \ \bigg\{q=(q_t)_{0\leq t\leq T}\colon\text{$q$ is a real-valued $\F$-adapted process such that $\E\int_0^T q_t^2\,dt<+\infty$}\bigg\}, \\
\Ac^{\xi,+} \ &= \ \bigg\{\xi\colon\Omega\rightarrow[0,+\infty)\colon\text{$\xi$ is an $\Fc_T$-measurable random variable}\bigg\}
\end{align*}
The expected total cost for agent $i$ is given by
\begin{equation}\label{Ji}
J_i(q^i,\xi_i) \ = \ \E\bigg[\int_0^T q_t^i\,\big(P_t + \gamma_i\,q_t^i\big)\,dt + c_i(\xi_i) + \frac{\eta_i}{2}\,(D_T^i - X_T^{i,q^i} - \xi_i)^2\bigg],
\end{equation}
where $\gamma_i$ and $\eta_i$ are positive constants, while $P$ denotes the intraday electricity quoted price, which will be endogenously determined in the following class of processes:

\vspace{2mm}

\noindent$\mathbf L^2(0,T)$ $=$ the set of all $\F$-adapted processes $P=(P_t)_{0\leq t\leq T}$ such that
\[
\E\bigg[\int_0^T|P_t|^2dt\bigg] \ < \ \infty.
\]

The agent's $i$ optimisation problem consist in trading at minimal cost to achieve a given terminal target, taking into account the liquidity cost of her sales or purchases. We take potentialy different impact parameter per agent $\gamma_i$, capturing here the potential different liquidity cost faced by market players. In this sense, we deviate from Almgren and Chriss (2001) \cite{Almgren01} and Aïd et. al. (2001) \cite{Aid16}, in the sense that there is no permanent market impact in agent's $i$ problem. Further, although in intraday electricity market, the same penalty cost is applied by the Transmission System Operator to any market player, we capture the idea that agents may have different appreciation of the cost of imbalances by using different imbalance cost parameter $\eta_i$. Thus, each agent $i$  is characterised by her cost function with Markov chain $\beta^i$, her valuation of imbalances $\eta_i$, her liquidity access $\gamma_i$, her demand forecast error function $\sigma^i$ and her correlation with the common noise $\rho_i$.

The optimization problem of agent $i$ consists in minimizing the expected total cost \eqref{Ji} over all admissible pairs of controls $(q,\xi)$ in $\Ac^q\times\Ac^{\xi,+}$. In order to solve such an optimization problem, we begin noting that we can easily find the optimal $\xi_i^{*,+}\in\Ac^{\xi,+}$ for agent $i$. As a matter of fact, in the expected total cost \eqref{Ji} the control $\xi_i$ appears only at the terminal time $T$. Then, the optimal $\xi_i^{*,+}$ is a non-negative $\Fc_T$-measurable random variable minimizing the quantity
\[
\E\bigg[c_i(\xi_i) + \frac{\eta_i}{2}\,\big(D_T^i - X_T^{i,q^i} - \xi_i\big)^2\bigg].
\]
It is then easy to see that $\xi_i^{*,+}$ is given by
\[
\xi_i^{*,+} \ = \ \frac{\eta_i}{\eta_i + \beta_T^i}\,\big(D_T^i - X_T^{i,q^i}\big)^+ \ = \ 
\begin{cases}
\dfrac{\eta_i}{\eta_i+\beta_T^i}\,\big(D_T^i-X_T^{i,q^i}\big), \qquad\qquad & D_T^i \geq X_T^{i,q^i}, \\
0, & D_T^i<X_T^{i,q^i}.
\end{cases}
\]

\subsection{Auxiliary optimal execution problem}
\label{SubS:Aux}

In the present section, inspired by \cite{Aid16}, we consider a relaxed version of the optimization problem for agent $i$, where the control $\xi_i$ is not constrained to be nonnegative, but it belongs to the set $\Ac^\xi$ defined as
\[
\Ac^\xi \ = \ \Big\{\xi\colon\Omega\rightarrow\R\colon\text{$\xi$ is an $\Fc_T$-measurable random variable}\Big\}.
\]
The optimization problem of agent $i$ now consists in minimizing the expected total cost \eqref{Ji} over all admissible pairs of controls $(q,\xi)$ in $\Ac^q\times\Ac^\xi$. From the expression of $J_i$ in \eqref{Ji}, it is straightforward to see that the optimal control $\xi_i^*$ is given by:
\[
\xi_i^* \ = \ \frac{\eta_i}{\eta_i + \beta_T^i}\,\big(D_T^i - X_T^{i,q^i}\big).
\]
Plugging $\xi_i^*$ into $J_i$, we find (to alleviate notation, we still denote by $J_i$ the new expected total cost, that now depends only on the control $q^i$)
\begin{equation}\label{Jiqi}
J_i(q^i) \ := \ J_i(q^i,\xi_i^*) \ = \ \E\bigg[\int_0^T q_t^i\,\big(P_t + \gamma_i\,q_t^i\big)\,dt + \frac{1}{2}\,\frac{\eta_i\,\beta_T^i}{\eta_i + \beta_T^i}\,\big(D_T^i - X_T^{i,q^i}\big)^2\bigg].
\end{equation}
In conclusion, the optimization problem of agent $i$ consists in minimizing \eqref{Jiqi} over all controls $q^i\in\Ac^q$. Because of the presence of the stochastic process $P$, we cannot solve such an optimization problem by means of the Bellman optimality principle, and, in particular, via PDE methods. For this reason, we rely on the martingale optimality principle, which can be implemented using only probabilistic techniques, based in particular on the theory of backward stochastic differential equations. More specifically, we solve the optimization problem of every agent finding $N$ optimal trading rates $\hat q^{1,P},\ldots,\hat q^{N,P}$, which depend on the price process $P$. Given the exogenous demands $(D^i)_i$ and production capacities  $(\beta^i)_i$, the equilibrium price $\hat P=(\hat P_t)_{0\leq t\leq T}$ is then obtained imposing the \emph{equilibrium condition}
\begin{equation}\label{Equilibrium}
\sum_{i=1}^N \hat q_t^{i,\hat P} \ = \ 0, \qquad \text{for all }0\leq t\leq T.
\end{equation}

\begin{Remark}{\rm Despite the homogeneous description of market players, the market model above allows to take into account a diversity of agents like pure retailers, pure producers or pure traders. Pure retailers have uncertain terminal demand $D^i_T$ but no generation plant. They can be represented taking a constant Markov chain $\beta^i$ taking a large value $e^i$. Pure producers have no demand $D^i_T$ to satisfy and are represented by the Markov chain of their generation cost. Finally, pure traders have neither a demand to satisfy nor generation asset, but only an initial inventory position.}
\end{Remark}

\section{Martingale optimality principle and optimal trading rates}
\label{S:Martingale}

The aim of this section is to find an optimal trading rate $\hat q^{i,P}$ of agent $i$ for every fixed price process $P$. In order to do it in the present non-Markovian framework (the non-Markovian feature is due to the presence of the process $P$), we consider a value process $V^{i,q_i}=(V_t^{i,q_i})_{0\leq t\leq T}$ given by
\begin{equation}\label{V}
V_t^{i,q^i} \ = \ \int_0^t q_s^i\,(P_s + \gamma_i\,q_s^i)ds + \big(D_t^i - X_t^{i,q^i}\big)^2\,Y_t^{2,i} + \big(D_t^i - X_t^{i,q^i}\big)\,Y_t^{1,i} + Y_t^{0,i},
\end{equation}
for all $0\leq t\leq T$, with $Y^{2,i}$, $Y^{1,i}$, $Y^{0,i}$ satisfying suitable backward stochastic differential equations, namely \eqref{hatY2}, \eqref{hatY1}, \eqref{hatY0} below. Then, the optimal trading rate $\hat q^{i,P}$ is obtained using the martingale optimality principle, namely imposing that for such a $\hat q^{i,P}$ the value process $V^{i,\hat q^{i,P}}$ is a true martingale, while it has to be a submartingale for any other trading rate $q^i$ (for more details on the martingale optimality principle see items (i)-(ii)-(iii) in the proof of Theorem \ref{Thm:Main1} below).

The present section is organized as follows. We firstly consider the three building blocks of formula \eqref{V}, namely equations \eqref{hatY2}, \eqref{hatY1}, \eqref{hatY0} (whose forms are chosen in order to satisfy the martingale/submartingale requirements of the value process) and prove an existence and uniqueness result for each of them. Then, exploiting the properties of the value process $V^{i,q^i}$, we prove the main result of this section, namely Theorem \ref{Thm:Main1}.

\subsection{Notations and preliminary results}

First of all, we introduce some notations. We denote by $\pi^i$ the jump measure of the Markov chain $\beta^i$, which is given by $\pi^i=\sum_{t\colon\beta_t^i\neq\beta_{t-}^i}\delta_{(t,\beta_t^i)}$, where $\delta_{(t,\beta_t^i)}$ is the Dirac delta at $(t,\beta_t^i)$. We also denote by $\nu^i$ the compensator of $\pi^i$, which has the following form (see for instance Section 8.3 and, in particular, Theorem 8.4 in \cite{DarlingNorris}):
\[
\nu^i(dt,\{e\}) \ = \ \lambda_i(\beta_{t-}^i,e)\,1_{\{\beta_{t-}^i\neq e\}}\,dt, \qquad \forall\,e\in E.
\]
In addition to the set $\mathbf L^2(0,T)$ previously defined, we introduce the following sets:
\begin{itemize}
\item $\mathbf S^\infty(0,T)$: \emph{the set of all bounded c\`adl\`ag $\F$-adapted processes on $[0,T]$.}
\item $\mathbf S^2(0,T)$: \emph{the set of all c\`adl\`ag $\F$-adapted processes $Y=(Y_t)_{0\leq t\leq T}$ satisfying \\ $\E\big[\sup_{0\leq t\leq T}|Y_t|^2\big]<\infty$.}
\item $\mathbf L_{\textup{Pred}}^2(0,T)$: \emph{the set of all $\F$-predictable processes $Z=(Z_t)_{0\leq t\leq T}$ satisfying $\E\big[\int_0^T|Z_t|^2dt\big]<\infty$.}
\item $\mathbf L_{\beta^i}^2(0,T)$: \emph{the set of all $\Pc\otimes\Bc(E)$-measurable maps $U\colon\Omega\times[0,T]\times E\rightarrow\R$ satisfying}  
\begin{align*}
\E\big[ \int_0^T \sum_{e\in E} |U_t(e)|^2 \lambda_i(\beta_{t-}^i,e) 1_{ \{ \beta_{t-}^i \neq e \} } dt \big] & <  \infty. 
\end{align*}
\emph{Here $\Bc(E)$ denotes the Borel $\sigma$-algebra of $E$, which turns out to be equal to the power set of $E$, since $E$ is a finite subset of $(0,+\infty)$.}
\end{itemize}

\paragraph{Construction of $Y^{2,i}$.} Let us construct the first building block of formula \eqref{V}, namely $Y^{2,i}$. First of all, for every $i=1,\ldots,N$, consider the following system of $M$ (recall that the set $E$ has cardinality $M$) coupled ordinary differential equations of Riccati type on the time interval $[0,T]$:
\begin{equation}\label{Riccati}
y_{i,\bar e}'(t) \ = \ \frac{1}{\gamma_i}|y_{i,\bar e}(t)|^2 - \sum_{e\in E} y_{i,e}(t)\lambda_i({\bar e},e), \qquad\qquad y_{i,\bar e}(T) \ = \ \frac{1}{2}\,\frac{\eta_i\,\bar e}{\eta_i + \bar e},
\end{equation}
for every $\bar e\in E$.

\begin{Lemma}\label{L:ODE}
For every $i=1,\ldots,N$, there exists a unique continuously differentiable solution 
$\mathbf y_i=(y_{i,e})_{e\in E}$ $\colon[0,T]\rightarrow\R^M$ to the system of equations \eqref{Riccati}. Moreover, every component $y_{i,e}$ of $\mathbf y_i$ is non-negative on the entire interval $[0,T]$.
\end{Lemma}
\textbf{Proof.}
For simplicity of notation, we fix $i\in\{1,\ldots,N\}$ and denote $\mathbf y_i=(y_{i,e})_{e\in E}$ simply by $\mathbf y=(y_e)_{e\in E}$. Notice that system \eqref{Riccati} can be equivalently rewritten in forward form as follows:
\begin{equation}\label{RiccatiForward}
\hat y_{\bar e}'(t) \ = \ - \frac{1}{\gamma_i}|\hat y_{\bar e}(t)|^2 + \sum_{e\in E} \hat y_e(t)\lambda_i({\bar e},e), \qquad\qquad \hat y_{\bar e}(0) \ = \ \frac{1}{2}\,\frac{\eta_i\,\bar e}{\eta_i + \bar e},
\end{equation}
with $\hat y_e(t)=y_e(T-t)$, for all $0\leq t\leq T$. By the classical Picard-Lindel\"of theorem (see for instance Theorem II.1.1 in \cite{Hartman}), it follows that there exists an interval $[0,\delta)\subset[0,+\infty)$ on which system \eqref{Riccati} admits a unique solution denoted by $\hat{\mathbf y}=(\hat y_e)_{e\in E}$. Let us prove that such a solution can be extended to the entire interval $[0,+\infty)$, so that, in particular, $\hat{\mathbf y}$ is defined on $[0,T]$.

According to standard extension theorems for ordinary differential equations (see for instance Corollary II.3.1 in \cite{Hartman}), it is enough to prove that the solution $\hat{\mathbf y}$ does not blow up in finite time. This holds true for system \eqref{RiccatiForward} as a consequence of the two following properties:
\begin{enumerate}
\item[1)] every component $\hat y_e$ of $\hat{\mathbf y}$ is non-negative on the entire interval $[0,+\infty)$;
\item[2)] the sum $\sum_{e\in E}\hat y_e'(t)$ is bounded from above by a constant independent of $t\in[0,+\infty)$.
\end{enumerate}
We begin proving item 1). Define $t_0=\inf\{t\geq0\colon\min_{e\in E}\hat y_e(t)\leq0\}$, with $\inf{\emptyset}=+\infty$. We prove that every $\hat y_e$, $e\in E$, is strictly positive on $[0,t_0)$ and identically equal to zero on $[t_0,+\infty)$ (in the case $t_0=+\infty$, every $\hat y_e$ is strictly positive on the entire interval $[0,\infty)$). If $t_0=+\infty$ there is nothing to prove. Therefore, suppose that $t_0<+\infty$, so that there exists $e_0\in E$ such that $\hat y_{e_0}(t_0)=0$. Since for every $e\in E$ we have $\hat y_e(0)>0$, then $t_0>0$ and, by continuity, every component $\hat y_e$ is strictly positive on the interval $[0,t_0)$. It remains to prove that every $\hat y_e$ is identically equal to zero on $[t_0,+\infty)$. Using equation \eqref{RiccatiForward}, this latter property follows if we prove that every $\hat y_e$ is equal to zero at $t_0$ (as a matter of fact, if this is true, then from equation \eqref{RiccatiForward} we deduce that every $\hat y_e$ remains at zero for all $t>t_0$). In order to prove that every $y_e$ is zero at $t_0$, we proceed by contradiction and assume that there exists $e_1\in E$ such that $\hat y_{e_1}(t_0)>0$. Then, it follows from equation \eqref{RiccatiForward} that $\hat y_{e_0}'(t_0)>0$. This is in contradiction with the fact that $\hat y_{e_0}$ is strictly positive on $[0,t_0)$ (which implies that $\hat y_{e_0}'(t_0)\leq0$). This concludes the proof of item 1).

Let us now prove item 2). Taking the sum over $\bar e\in E$ in equation \eqref{RiccatiForward}, we obtain
\begin{equation}\label{y'Bound}
\sum_{\bar e\in E} \hat y_{\bar e}'(t) \ = \ - \frac{1}{\gamma_i}\sum_{\bar e\in E}|\hat y_{\bar e}(t)|^2 + \sum_{\bar e,e\in E} \hat y_e(t)\lambda_i({\bar e},e).
\end{equation}
By Young's inequality ($ab\leq a^2/(2\gamma_i) + \gamma_i\,b^2/2$) we find
\begin{equation}\label{y'Bound2}
\sum_{\bar e,e\in E} \hat y_e(t)\lambda_i({\bar e},e) \ = \ \sum_{e\in E} \hat y_e(t) \bigg(\sum_{\bar e\in E}\lambda_i({\bar e},e)\bigg) \ \leq \ \frac{1}{2\gamma_i}\sum_{e\in E} |\hat y_e(t)|^2 + \hat C,
\end{equation}
with
\[
\hat C \ := \ \frac{\gamma_i}{2}\sum_{e\in E}\bigg|\sum_{\bar e\in E}\lambda_i({\bar e},e)\bigg|^2.
\]
Plugging \eqref{y'Bound2} into \eqref{y'Bound}, we end up with
\[
\sum_{\bar e\in E} \hat y_{\bar e}'(t) \ \leq \ - \frac{1}{2\gamma_i}\sum_{\bar e\in E}|\hat y_{\bar e}(t)|^2 + \hat C \ \leq \ \hat C,
\]
which concludes the proof of item 2).
\ep

\vspace{3mm}

By Lemma \ref{L:ODE}, we know that there exists a unique $C^1$-solution $\mathbf y_i=(y_{i,e})_{e\in E}$ to system \eqref{Riccati}. Then, define the stochastic process
\begin{equation}\label{BSDE_Riccati}
Y_t^{2,i} \ = \ y_{i,\beta_t^i}(t), \qquad \text{for all }0\leq t\leq T.
\end{equation}
As it will be proved in Proposition \ref{P:BSDE1} below, $Y^{2,i}$ solves the following backward stochastic differential equation on $[0,T]$, driven by the Markov chain $\beta^i$, with quadratic growth in the component $Y^{2,i}$:
\begin{equation}\label{hatY2}
Y_t^{2,i} \ = \ \frac{1}{2}\,\frac{\eta_i\,\beta_T^i}{\eta_i + \beta_T^i} + \int_t^T f_s^{2,i}\,ds - \int_{(t,T]\times E} U_s^{2,i}(e)\,(\pi^i-\nu^i)(ds,de),
\end{equation}
for all $0\leq t\leq T$, where
\begin{equation}\label{fhati,2}
f_t^{2,i} \ = \ -\frac{1}{\gamma_i}|Y_t^{2,i}|^2
\end{equation}
and
\begin{equation}\label{Uhati,2}
U_t^{2,i}(e) \ = \ y_{i,e}(t) - y_{i,\beta_{t-}^i}(t).
\end{equation}
\begin{Proposition}\label{P:BSDE1}
For every $i=1,\ldots,N$, the backward equation \eqref{hatY2} admits a unique solution $(Y^{2,i},U^{2,i})\in\mathbf S^\infty(0,T)\times\mathbf L_{\beta^i}^2(0,T)$ given by \eqref{BSDE_Riccati} and \eqref{Uhati,2}. Moreover, $Y^{2,i}$ is non-negative.
\end{Proposition}
\textbf{Proof.} Let $(Y^{2,i},U^{2,i})$ be the pair given by \eqref{BSDE_Riccati} and \eqref{Uhati,2}. Notice that $(Y^{2,i},U^{2,i})$ belongs to $\mathbf S^\infty(0,T)\times\mathbf L_{\beta^i}^2(0,T)$. As a matter of fact
\[
\sup_{0\leq t\leq T}\big|Y_t^{2,i}\big| \ = \ \sup_{0\leq t\leq T}|y_{i,\beta_t^i}(t)| \ \leq \ \sup_{0\leq t\leq T} \max_{e\in E}|y_{i,e}(t)| \ < \ +\infty
\]
and
\begin{align*}
\E\bigg[\int_0^T\sum_{e\in E}\big|U_t^{2,i}(e)\big|^2\lambda_i(\beta_{t-}^i,e)dt\bigg] \ &= \ \E\bigg[\int_0^T\sum_{e\in E}|y_{i,e}(t) - y_{i,\beta_{t-}^i}(t)|^2\lambda_i(\beta_{t-}^i,e)dt\bigg] \\
&\leq \ \E\bigg[\int_0^T\sum_{e\in E}\big(2|y_{i,e}(t)|^2 + 2|y_{i,\beta_{t-}^i}(t)|^2\big)\lambda_i(\beta_{t-}^i,e)dt\bigg] \\
&\leq \ 4M\int_0^T\max_{e\in E}|y_{i,e}(t)|^2\Big(\sum_{e\in E}\lambda_i(\beta_{t-}^i,e)\Big)dt \ < \ +\infty,
\end{align*}
where recall that $M$ is the cardinality of the set $E$. It remains to prove that $(Y^{2,i},U^{2,i})$ solves equation \eqref{hatY2}. Applying It\^o's formula to $y_{i,\beta_\cdot^i}(\cdot)$ between $t\in[0,T)$ and $T$, we find
\begin{equation}\label{BSDE1_1}
y_{i,\beta_T^i}(T) \ = \ y_{i,\beta_t^i}(t) + \int_t^T y_{i,\beta_s^i}'(s)ds + \sum_{t<s\leq T}\big(y_{i,\beta_s^i}(s) - y_{i,\beta_{s-}^i}(s)\big).
\end{equation}
Now, since $\sum_{e\in E} \lambda_i({\bar e},e)=0$, equation \eqref{Riccati} can be rewritten as follows
\[
y_{i,\bar e}'(t) \ = \ \frac{1}{\gamma_i}|y_{i,\bar e}(t)|^2 - \sum_{e\in E} \big(y_{i,e}(t) - y_{i,\bar e}(t)\big)\lambda_i({\bar e},e).
\]
Therefore
\begin{equation}\label{BSDE1_2}
\int_t^T y_{i,\beta_s^i}'(s)ds \ = \ - \int_t^T \hat f_s^{2,i}\,ds - \int_t^T\int_E U_s^{2,i}(e)\,\nu^i(ds,de).
\end{equation}
On the other hand, we have
\begin{equation}\label{BSDE1_3}
\sum_{t<s\leq T}\big(y_{i,\beta_s^i}(s) - y_{i,\beta_{s-}^i}(s)\big) \ = \ \int_{(t,T]\times E} U_s^{2,i}(e)\,\pi^i(ds,de).
\end{equation}
Hence, plugging \eqref{BSDE1_2} and \eqref{BSDE1_3} into \eqref{BSDE1_1}, we obtain equation \eqref{hatY2}.
\ep

\paragraph{Construction of $Y^{1,i}$.} Let us construct the second ingredient of formula \eqref{V}, namely $Y^{1,i}$, which will be denoted by $Y^{1,i,P}$ to emphasize its dependence on $P$. For every $i=1,\ldots,N$ and any $P\in\mathbf L^2(0,T)$, consider the following linear backward stochastic differential equation on $[0,T]$, driven by the Brownian motions $W^0,W^1,\ldots,W^N$ and the Markov chains $\beta^1,\ldots,\beta^N$:
\begin{equation}\label{hatY1}
Y_t^{1,i,P} = \int_t^T f_s^{1,i,P} ds - \sum_{j=0}^N\int_t^T Z_s^{1,i,j,P} d W_s^j - \sum_{j=1}^N\int_{(t,T]\times E} U_s^{1,i,j,P}(e)\,(\pi^j-\nu^j)(ds,de),
\end{equation}
where
\begin{equation}\label{fhati,1}
f_t^{1,i,P} \ = \ 2\mu_t^i Y_t^{2,i} + \frac{1}{\gamma_i}Y_t^{2,i}\big(P_t - Y_t^{1,i,P}\big).
\end{equation}
Notice that equation \eqref{hatY1} has zero terminal condition at time $T$: $Y_T^{1,i,P}=0$. We also observe that the generator depends linearly on the component $Y^{1,i,P}$ and it is random (as it depends on $Y^{2,i}$ and $P$). We now address the problem of existence and uniqueness of a solution to equation \eqref{hatY1}, for which we need the following martingale representation result.

\begin{Lemma}\label{L:MartRepr}
For every square-integrable real-valued $\Fc_T$-measurable random variable $\zeta$, there exist $Z^0,Z^1,\ldots,Z^N\in\mathbf L_{\textup{Pred}}^2(0,T)$, $U^1\in\mathbf L_{\beta^1}^2(0,T)$, $\ldots$, $U^N\in\mathbf L_{\beta^N}^2(0,T)$ such that
\begin{equation}\label{MartRepr}
\zeta \ = \ \E[\zeta] + \sum_{j=0}^N\int_0^T Z_s^j d W_s^j + \sum_{j=1}^N\int_{(0,T]\times E} U_s^j(e)\,(\pi^j-\nu^j)(ds,de).
\end{equation}
\end{Lemma}
\textbf{Proof.}
The result is standard and follows for instance from Example 2.1-(2) in \cite{Becherer02}. For completeness, we report the main steps of the proof. Denote by $\F^W=(\Fc_t^W)_{t\geq0}$ (resp. $\F^{\beta^i}=(\Fc_t^{\beta^i})_{t\geq0}$) the augmentation of the filtration generated by $(W^0,W^1,\ldots,W^N)$ (resp. $\beta^i$). It is well-known that if $\zeta$ is $\Fc_T^W$-measurable (resp. $\Fc_T^{\beta^i}$-measurable) then representation \eqref{MartRepr} holds; indeed, in this case representation \eqref{MartRepr} is such that $U^1,\ldots,U^N$ (resp. $Z^0,Z^1,\ldots,Z^N$ and $U^j$, $j\neq i$) are equal to zero.

It is then easy to see that representation \eqref{MartRepr} also holds for every $\zeta$ of the form $\zeta_0\zeta_1\cdots\zeta_N$, with $\zeta_0$ and $\zeta_i$, $i\in\{1,\ldots,N\}$, being respectively $\Fc_T^W$-measurable and $\Fc_T^{\beta^i}$-measurable. The claim follows from the fact that the linear span of the random variables of the form $\zeta_0\zeta_1\cdots\zeta_N$ is dense in $L^2(\Omega,\Fc_T,\P;\R)$ (the space of square-integrable real-valued $\Fc_T$-measurable random variables).
\ep
 
\begin{Proposition}\label{P:FBSDE}
For every $i=1,\ldots,N$ and any $P\in\mathbf L^2(0,T)$, the backward equation \eqref{hatY1} admits a unique solution $(Y^{1,i,P},Z^{1,i,0,P},Z^{1,i,1,P},\ldots,Z^{i,1,N,P},U^{1,i,1,P},\ldots,U^{1,i,N,P})\in\mathbf S^2(0,T)\times\mathbf L_{\textup{Pred}}^2(0,T)\times\cdots\times\mathbf L_{\textup{Pred}}^2(0,T)\times\mathbf L_{\beta^1}^2(0,T)\times\cdots\times\mathbf L_{\beta^N}^2(0,T)$. Moreover, $Y^{1,i,P}$ is given by
\begin{equation}\label{yi1}
Y_t^{1,i,P} \ = \ \frac{1}{\Gamma_t^i}\E\bigg[\int_t^T \Gamma_s^i Y_s^{2,i}\Big(2\mu_i+\frac{1}{\gamma_i}P_s\Big)ds\bigg|\Fc_t\bigg], \qquad \P\text{-a.s.}
\end{equation}
for all $0\leq t\leq T$, where $\Gamma_t^i=e^{-\frac{1}{\gamma_i}\int_0^t  Y_s^{2,i}ds}=e^{-\frac{1}{\gamma_i}\int_0^t y_{i,\beta_s^i}(s)ds}$.
\end{Proposition}
\textbf{Proof.}
\emph{Existence.} Fix $i\in\{1,\ldots,N\}$, $P\in\mathbf L^2(0,T)$ and define (to alleviate notation, we write $\zeta^i$ rather than $\zeta^{i,P}$ as $P$ is fixed throughout the proof; we adopt the same convention for all the other quantities involved in the proof)
\[
\zeta^i \ = \ \int_0^T \Gamma_s^i Y_s^{2,i}\Big(2\mu_i+\frac{1}{\gamma_i}P_s\Big)ds.
\]
Since $\zeta^i$ is a square-integrable real-valued $\Fc_T$-measurable random variable, we can apply Lemma \ref{L:MartRepr} from which we deduce the existence of $\hat Z^{1,i,0},\hat Z^{1,i,1},\ldots,\hat Z^{1,i,N}\in\mathbf L_{\textup{Pred}}^2(0,T)$, $\hat U^{1,i,1}\in\mathbf L_{\beta^1}^2(0,T)$, $\ldots$, $\hat U^{1,i,N}\in\mathbf L_{\beta^N}^2(0,T)$ such that
\begin{equation}\label{MartReprXi}
\zeta^i \ = \ \E[\zeta^i] + \sum_{j=0}^N\int_0^T \hat Z_s^{1,i,j} d W_s^j + \sum_{j=1}^N\int_{(0,T]\times E} \hat U_s^{1,i,j}(e)\,(\pi^j-\nu^j)(ds,de).
\end{equation}
Now, define $\hat Y^{1,i}=(\hat Y_t^{1,i})_{0\leq t\leq T}$ as (the c\`adl\`ag version of)
\[
\bigg(\E\bigg[\int_t^T \Gamma_s^i Y_s^{2,i}\Big(2\mu_i+\frac{1}{\gamma_i}P_s\Big)ds\bigg|\Fc_t\bigg]\bigg)_{0\leq t\leq T}.
\]
Since $P\in\mathbf L^2(0,T)$, we see that $\hat Y^{1,i}\in\mathbf S^2(0,T)$. Moreover, taking the conditional expectation with respect to $\Fc_t$ in \eqref{MartReprXi}, we obtain
\[
\hat Y_t^{1,i} \ = \ \hat Y_0^{1,i} - \int_0^t \Gamma_s^i Y_s^{2,i}\Big(2\mu_i+\frac{1}{\gamma_i}P_s\Big) ds + \sum_{j=0}^N\int_0^t \hat Z_s^{1,i,j} d W_s^j + \sum_{j=1}^N\int_{(0,t]\times E} \hat U_s^{1,i,j}(e)\,(\pi^j-\nu^j)(ds,de).
\]
Finally, we define $Y^{1,i}=(Y_t^{1,i})_{0\leq t\leq T}$ as $Y_t^{1,i}=\hat Y_t^{1,i}/\Gamma_t^i$. Then, noting that
\[
d\Gamma_t^i \ = \ -\frac{1}{\gamma_i}Y_t^{2,i}\Gamma_t^i dt, \qquad\qquad\qquad\qquad \Gamma_0^i \ = \ 1,
\]
applying It\^o's formula to $\hat Y_t^{1,i}/\Gamma_t^i$, we get
\begin{align*}
Y_t^{1,i} \ &= \ Y_0^{1,i} - \int_0^t Y_s^{2,i}\Big(2\mu_i+\frac{1}{\gamma_i}P_s\Big)ds + \sum_{j=0}^N\int_t^T Z_s^{1,i,j} d W_s^j \\
&\quad \ + \sum_{j=1}^N\int_{(t,T]\times E} U_s^{1,i,j}(e)\,(\pi^j-\nu^j)(ds,de) + \frac{1}{\gamma_i}\int_0^t Y_s^{1,i}Y_s^{2,i}ds,
\end{align*}
where
\[
Z_t^{1,i,j} \ = \ \frac{\hat Z_t^{1,i,j}}{\Gamma_t^i}, \qquad\qquad U_t^{1,i,j}(e) \ = \ \frac{\hat U_t^{1,i,j}(e)}{\Gamma_t^i}.
\]
This proves that $(Y^{1,i},Z^{1,i,0},Z^{1,i,1},\ldots,Z^{1,i,N},U^{1,i,1},\ldots,U^{1,i,N})$ solves equation \eqref{hatY1}; moreover, since $(\Gamma^i)^{-1}\in\mathbf S^\infty(0,T)$, it easy to see that such a solution belongs to $\mathbf S^2(0,T)\times\mathbf L_{\textup{Pred}}^2(0,T)\times\cdots\times\mathbf L_{\textup{Pred}}^2(0,T)\times\mathbf L_{\beta^1}^2(0,T)\times\cdots\times\mathbf L_{\beta^N}^2(0,T)$.

\vspace{1mm}

\noindent\emph{Uniqueness.} Fix $i\in\{1,\ldots,N\}$ and let $(\tilde Y^{1,i},\tilde Z^{1,i,0},\tilde Z^{1,i,1},\ldots,\tilde Z^{1,i,N},\tilde U^{1,i,1},\ldots,\tilde U^{1,i,N})\in\mathbf S^2(0,T)\times\mathbf L_{\textup{Pred}}^2(0,T)\times\cdots\times\mathbf L_{\textup{Pred}}^2(0,T)\times\mathbf L_{\beta^1}^2(0,T)\times\cdots\times\mathbf L_{\beta^N}^2(0,T)$ be a solution to equation \eqref{hatY1}. Applying It\^o's formula to the product $\Gamma_t^i\tilde Y_t^{1,i}$, it is easy to see that $\tilde Y^{1,i}$ is given by \eqref{yi1}. This proves the uniqueness of the $Y$-component, which in turn implies the uniqueness of all other components and concludes the proof.
\ep

\paragraph{Construction of $Y^{0,i}$.} Let us finally construct the third and last ingredient of formula \eqref{V}, namely $Y^{0,i,P}$, which will be denoted by $Y^{0,i,P}$ to emphasize its dependence on $P$. For every $i=1,\ldots,N$ and any $P\in\mathbf L^2(0,T)$, consider the following backward stochastic differential equation on $[0,T]$, driven by the Brownian motions $W^0,W^1,\ldots,W^N$ and the Markov chains $\beta^1,\ldots,\beta^N$:
\begin{equation}\label{hatY0}
Y_t^{0,i,P} = \int_t^T f_s^{0,i,P} ds - \sum_{j=0}^N\int_t^T Z_s^{0,i,j,P} dW_s^j - \sum_{j=1}^N\int_{(t,T]\times E} U_s^{0,i,j,P}(e)\,(\pi^j-\nu^j)(ds,de),
\end{equation}
where
\begin{equation}\label{fhati,0}
f_t^{0,i,P} \ := \ |\sigma_t^i|^2 Y_t^{2,i} + \mu_i Y_t^{1,i,P} + \sigma_t^i\rho_i Z_t^{1,i,0,P} + \sigma_t^i\sqrt{1-\rho_i^2}Z_t^{1,i,i,P} - \frac{1}{4\gamma_i}\big(P_t -  Y_t^{1,i,P}\big)^2.
\end{equation}
Notice that equation \eqref{hatY0} has zero terminal condition at time $T$: $Y_T^{0,i,P}=0$.

\begin{Proposition}\label{P:FBSDE0}
For every $i=1,\ldots,N$ and any $P\in\mathbf L^2(0,T)$, the backward equation \eqref{hatY0} admits a unique solution $(Y^{0,i,P},Z^{0,i,0,P},Z^{0,i,1,P},\ldots,Z^{0,i,N,P},U^{0,i,1,P},\ldots,U^{0,i,N,P})\in\mathbf S^2(0,T)\times\mathbf L_{\textup{Pred}}^2(0,T)\times\cdots\times\mathbf L_{\textup{Pred}}^2(0,T)\times\mathbf L_{\beta^1}^2(0,T)\times\cdots\times\mathbf L_{\beta^N}^2(0,T)$. Moreover, $Y^{0,i,P}$ is given by
\[
Y_t^{0,i,P} \ = \ \E\bigg[\int_t^T \Big(|\sigma_s^i|^2 Y_s^{2,i} + \mu_i Y_s^{1,i,P} + \sigma_s^i\rho_i Z_s^{1,i,0,P} + \sigma_s^i\sqrt{1-\rho_i^2}Z_s^{1,i,i,P} - \frac{1}{4\gamma_i}\big(P_s -  Y_s^{1,i,P}\big)^2\Big)ds\bigg|\Fc_t\bigg], \quad \P\text{-a.s.}
\]
for all $0\leq t\leq T$.
\end{Proposition}
\textbf{Proof.}
The result can be proved proceeding along the same lines as in the proof of Proposition \ref{P:FBSDE}, noting that the backward equation is still linear (in this case, the generator does not even depend on the unknowns).
\ep

\subsection{Main result}

We can finally state our main result, which provides the optimal trading rate of agent $i$ given a fixed price process $P$.

\begin{Theorem}\label{Thm:Main1}
For every $i=1,\ldots,N$ and any $P\in\mathbf L^2(0,T)$, there exists a unique (up to $\P$-indistinguishability) continuous process $\hat X^{i,P}=(\hat X_t^{i,P})_{0\leq t\leq T}$ in $\mathbf L_{\textup{Pred}}^2(0,T)$ satisfying the following equation:
\begin{equation}\label{X}
\hat X_t^{i,P} \ = \ x_0^i + \frac{1}{2\gamma_i} \int_0^t \Big(2 Y_s^{2,i}\big(D_s^i - \hat X_s^{i,P}\big) + Y_s^{1,i,P} - P_s\Big)\,ds, \qquad \text{for all }0\leq t\leq T,\;\P\text{-a.s.}
\end{equation}
with $Y^{2,i}$ and $Y^{1,i,P}$ given respectively by \eqref{BSDE_Riccati} and \eqref{yi1}. Define
\begin{equation}\label{qhat}
\hat q_t^{i,P} \ = \ \frac{1}{2\gamma_i} \Big(2 Y_t^{2,i}\big(D_t^i - \hat X_t^{i,P}\big) + Y_t^{1,i,P} - P_t\Big), \qquad \text{for all }0\leq t\leq T. 
\end{equation}
Then $\hat X^{i,P}\equiv X^{i,\hat q^{i,P}}$ and the following holds: 
\begin{itemize}
\item[\textup{1)}] $\hat q^{i,P}$ is an admissible control: $\hat q^{i,P}\in\Ac^q$;
\item[\textup{2)}] $\hat q^{i,P}$ is an optimal control for agent $i$.
\end{itemize}
\end{Theorem}
\textbf{Proof.}
Concerning equation \eqref{X}, notice that such an equation is deterministic with stochastic coefficients, so it can be solved pathwise. More precisely, \eqref{X} is a first-order linear ordinary differential equation (with stochastic coefficients), so that it admits a unique solution which can be written in explicit form. It is then clear that such a solution is continuous and $\F$-adapted, since all the coefficients are also $\F$-adapted.

It remains to prove items 1) and 2). To this end, fix $i=1,\ldots,N$ and $P\in\mathbf L^2(0,T)$ (to alleviate notation, in the sequel we do not explicitly report the dependence on $P$; so, for instance, we simply write $\hat X^i,Y^{1,i},\hat q^i$ instead of $\hat X^{i,P},Y^{1,i,P},\hat q^{i,P}$). The admissibility of $\hat q^i$ follows directly from its definition, using the integrability properties of $P$, $D^i$, $\hat X^i$, $Y^{1,i}$, $Y^{2,i}$. Let us now prove item 2). In order to prove the optimality of $\hat q^i$, we implement the martingale optimality principle. More precisely, we construct a family of processes $(V_t^{i,q^i})_{0\leq t\leq T}$, for every $q^i\in\Ac^q$, satisfying the following properties:
\begin{itemize}
\item[(i)] for every $q^i\in\Ac^q$, we have
\[
V_T^{i,q^i} \ = \ \int_0^T q_t^i\,\big(P_t + \gamma_i\,q_t^i\big)\,dt +  \frac{1}{2}\frac{\eta_i\,\beta_T^i}{\eta_i + \beta_T^i}\,\big(D_T^i - X_T^{i,q^i}\big)^2.
\]
\item[(ii)] $V_0^{i,q^i}$ is a constant independent of $q^i\in\Ac^q$.
\item[(iii)] $V^{i,q^i}$ is a submartingale for all $q^i\in\Ac^q$, and $V^{i,\hat q^i}$ is a martingale when $q^i=\hat q^i$.
\end{itemize}
Notice that when $q^i$ is given by $\hat q^i$ then $X^{i,\hat q^i}\equiv\hat X^i$, with $\hat X^i$ satisfying \eqref{X}. Suppose for a moment that we have already constructed a family of stochastic processes $(V_t^{i,q^i})_{0\leq t\leq T}$, $q^i\in\Ac^q$, satisfying points (i)-(ii)-(iii). Then, observe that, for any $q^i\in\Ac^q$, we have
\begin{align*}
J_i(q^i) \ = \ \E\bigg[\int_0^T q_t^i\,\big(P_t + \gamma_i\, q_t^i\big)\,dt + \frac{1}{2}\frac{\eta_i\,\beta_T^i}{\eta_i + \beta_T^i}\,\big(D_T^i - X_T^{i,q^i}\big)^2\bigg] \ = \ \E\big[V_T^{i,q^i}\big] \ \geq \ V_0^{i,q^i} = V_0^{i,\hat q^i} \\
= \ \E\big[V_T^{i,\hat q^i}\big] \ = \ \E\bigg[\int_0^T \hat q_t^i\,\big(P_t + \gamma_i\,\hat q_t^i\big)\,dt + \frac{1}{2}\frac{\eta_i\,\beta_T^i}{\eta_i + \beta_T^i}\,\big(D_T^i - \hat X_T^i\big)^2\bigg] \ = \ J_i(\hat q^i),
\end{align*}
which proves the optimality of $\hat q^i$. It remains to construct $(V_t^{i,q^i})_{0\leq t\leq T}$, $q^i\in\Ac^q$, satisfying (i)-(ii)-(iii). Given $q^i\in\Ac^q$, we take $(V_t^{i,q^i})_{0\leq t\leq T}$ as in \eqref{V}, namely
\[
V_t^{i,q^i} \ = \ \int_0^t q_s^i\,(P_s + \gamma_i\,q_s^i)ds + \big(D_t^i - X_t^{i,q^i}\big)^2\,Y_t^{2,i} + \big(D_t^i - X_t^{i,q^i}\big)\,Y_t^{1,i} + Y_t^{0,i},
\]
for all $0\leq t\leq T$, with $Y^{2,i}$, $Y^{1,i}$, $Y^{0,i}$ satisfying respectively \eqref{hatY2}, \eqref{hatY1}, \eqref{hatY0}.

From the definition of $(V_t^{i,q^i})_{0\leq t\leq T}$, it is clear that (i) holds. Moreover, since $\hat Y^{2,i}$, $\hat Y^{1,i}$, $\hat Y^{0,i}$ are independent of $q^i$, we see that (ii) holds as well. It remains to prove item (iii). By It\^o's formula we obtain $V_t^{i,q^i}=V_0^{i,q^i}+\int_0^t b_s^{i,q^i} ds + \text{martingale}$, where
\begin{align*}
b_t^{i,q^i} \ &= \ q_t^i(P_t + \gamma_i q_t^i) - \big(D_t^i - X_t^{i,q^i}\big)^2 f_t^{2,i} - \big(D_t^i - X_t^{i,q^i}\big) f_t^{1,i} - f_t^{0,i} \\
&\quad \ + Y_t^{2,i} \big[2\big(D_t^i - X_t^{i,q^i}\big)(\mu_i - q_t^i) + |\sigma_t^i|^2\big] + (\mu_i - q_t^i)Y_t^{1,i} + \sigma_t^i\rho_i Z_t^{1,i,0} + \sigma_t^i\sqrt{1-\rho_i^2}Z_t^{1,i,i}.
\end{align*}
It is easy to see that when $q^i=\hat q^i$ the drift $b^{i,\hat q^i}$ becomes zero. So, in particular, $V^{i,\hat q^i}$ is a true martingale. In order to conclude the proof, we need to prove that in general we have $b^{i,q^i}\geq0$, that is $(V_t^{i,q^i})_{0\leq t\leq T}$ is a submartingale for any $q^i$. To this end, it is useful to rewrite $b_t^{i,q^i}$ as a quadratic polynomial in the variable $q_t^i$:
\begin{align*}
b_t^{i,q^i} \ &= \ \gamma_i |q_t^i|^2 + \big[P_t - 2 Y_t^{2,i}\big(D_t^i - X_t^{i,q^i}\big) - Y_t^{1,i}\big]q_t^i - \big(D_t^i - X_t^{i,q^i}\big)^2 f_t^{2,i} \\
&\quad \ - \big(D_t^i - X_t^{i,q^i}\big) f_t^{1,i} - f_t^{0,i} + Y_t^{2,i} \big[2\big(D_t^i - X_t^{i,q^i}\big)\mu + |\sigma_t^i|^2\big] + \mu_i Y_t^{1,i} + \sigma_t^i\rho_i Z_t^{1,i,0} + \sigma_t^i\sqrt{1-\rho_i^2}Z_t^{1,i,i}.
\end{align*}
Since $\gamma_i>0$, $b_t^{i,q^i}$ is nonnegative for every value of $q_t^i$ if and only if the discriminant is nonpositive. Notice however that the discriminant cannot be strictly negative, otherwise this would give a contradiction to the fact that $b^{i,\hat q^i}$ is zero. In conclusion, the discriminant has be identically equal to zero, namely
\begin{align*}
4 \gamma_i \bigg\{-\big(D_t^i - X_t^{i,q^i}\big)^2 f_t^{2,i} - \big(D_t^i - X_t^{i,q^i}\big)f_t^{1,i} - f_t^{0,i} + Y_t^{2,i} \big[2\big(D_t^i - X_t^{i,q^i}\big)\mu_i + |\sigma_t^i|^2&\big] \\
+\,\mu_i Y_t^{1,i} + \sigma_t^i\rho_i Z_t^{1,i,0} + \sigma_t^i\sqrt{1-\rho_i^2}Z_t^{1,i,i}\bigg\} \ = \ \big[P_t - 2 Y_t^{2,i} \big(D_t^i - X_t^{i,q^i}\big) - Y_t^{1,i}&\big]^2.
\end{align*}
Rewriting it in terms of the variable $D_t^i-X_t^{i,q^i}$, we find
\begin{align*}
4\big(- \gamma_i f_t^{2,i} - |Y_t^{2,i}|^2\big)\big(D_t^i - X_t^{i,q^i}\big)^2 + 4\big[\gamma_i\big(2\mu_i Y_t^{2,i} - f_t^{1,i}\big) + Y_t^{2,i}(P_t - Y_t^{1,i})\big]\big(D_t^i - X_t^{i,q^i}\big) \\
+\,4 \gamma_i\big(|\sigma_t^i|^2 Y_t^{2,i} + \mu_i Y_t^{1,i} + \sigma_t^i\rho_i Z_t^{1,i,0} + \sigma_t^i\sqrt{1-\rho_i^2}Z_t^{1,i,i} - f_t^{0,i}\big) - (P_t - Y_t^{1,i})^2 \ = \ 0.
\end{align*}
Now, we see that $f^{2,i}$, $f^{1,i}$, $f^{0,i}$ (defined in \eqref{fhati,2}, \eqref{fhati,1}, \eqref{fhati,0}, respectively) are such that the above equality is always satisfied, regardless of the value of $D_t^i-X_t^{i,q^i}$. It follows that $b^{i,q^i}$ is nonnegative, which implies that $(V_t^{i,q^i})_{0\leq t\leq T}$ is a submartingale and concludes the proof.
\ep

\section{Equilibrium price}
\label{S:Equilibrium}

In the present section we use the explicit expression of $\hat q^{i,P}$ in \eqref{qhat} together with the equilibrium condition \eqref{Equilibrium} to find the equilibrium price process $\hat P=(\hat P_t)_{0\leq t\leq T}$ (Theorem \ref{Thm:Main2}). We also find the dynamics of the equilibrium price process (Theorem \ref{Thm:P_Dynamics}), and obtain notably the martingale property of the equilibrium price process.

\begin{Theorem}\label{Thm:Main2}
There exists a unique solution $(\hat X^i,\hat Y^{1,i},\hat Z^{1,i,0},\hat Z^{1,i,j},\hat U^{1,i,j})_{i,j=1,\ldots,N}$, with $\hat X^i\in\mathbf L_{\textup{Pred}}^2(0,T)$ being a continuous process, $\hat Y^{1,i}\in\mathbf S^2(0,T)$, $\hat Z^{1,i,j}\in\mathbf L_{\textup{Pred}}^2(0,T)$, $\hat U^{1,i,j}\in\mathbf L_{\beta^j}^2(0,T)$, satisfying the following coupled forward-backward system of stochastic differential equations:
\begin{numcases}
\,\hat X_t^i \ \, \displaystyle= \ x_0^i + \frac{1}{2\gamma_i} \int_0^t \big(2 Y_s^{2,i}\big(D_s^i - \hat X_s^i\big) + \hat Y_s^{1,i} - \hat P_s\big)\,ds, \quad\; 0\leq t\leq T, \label{CoupledFBSDEX} \\
\hat Y_t^{1,i} \displaystyle= \ \int_t^T \Big(2\mu_i Y_s^{2,i} + \frac{1}{\gamma_i}Y_s^{2,i}\big(\hat P_s - \hat Y_s^{1,i}\big)\Big) ds - \sum_{j=0}^N\int_t^T \hat Z_s^{1,i,j} d W_s^j \label{CoupledFBSDEY} \\
\displaystyle\qquad \quad\; - \sum_{j=1}^N\int_{(t,T]\times E} \hat U_s^{1,i,j}(e)\,(\pi^j-\nu^j)(ds,de), \qquad\quad\;\;\, 0\leq t\leq T, \notag
\end{numcases}
where
\begin{equation}
\hat P_t \ := \ \sum_{j=1}^N  \frac{\bar\gamma}{\gamma_j}\Big(  2 Y_t^{2,j}\big(D_t^j - \hat X_t^j\big) + \hat Y_t^{1,j} \Big), \qquad \bar \gamma \ :=  \ \bigg(\sum_{i=1}^N \frac1\gamma_i\bigg)^{-1}, \qquad \text{for all }0\leq t\leq T. \label{Phat}
\end{equation}
Moreover, $\hat X^i$ coincides with $\hat X^{i,\hat P}$ of equation \eqref{X}, while $\hat Y^{1,i},\hat Z^{1,i,j},\hat U^{1,i,j}$ coincide with $\hat Y^{1,i,\hat P},\hat Z^{1,i,j,\hat P},\hat U^{1,i,j,\hat P}$ of equation \eqref{hatY1}. Finally, $\hat P$ is the price process satisfying the equilibrium condition \eqref{Equilibrium} with $\hat q^{i,\hat P}$ as in \eqref{qhat}.
\end{Theorem}
\textbf{Proof.}
Existence and uniqueness for system \eqref{CoupledFBSDEX}-\eqref{CoupledFBSDEY} can be proved proceeding along the same lines as in the proof of Lemma 2.2 in \cite{LiWei}, the only difference being that $\pi^j$ is a Poisson random measure in \cite{LiWei}. We also notice that, proceeding as in the proof of Proposition 3.1 in \cite{LiWei}, we obtain the following estimate:
\begin{align}\label{Estimate-coupled-system}
\sum_{i=1}^N\E\bigg[\sup_{0\leq t\leq T}|\hat X_t^i|^2 + \sup_{0\leq t\leq T}|\hat Y_t^{1,i}|^2 &+ \sum_{j=0}^N\int_0^T |\hat Z_t^{1,i,j}|^2\,dt \notag \\
&+ \sum_{j=1}^N\int_0^T \sum_{e\in E} |\hat U_t^{1,i,j}(e)|^2\,\lambda_i(\beta_{t-}^i,e)\,1_{\{\beta_{t-}^i\neq e\}}\,dt \bigg] \ \leq \ \hat C,	
\end{align}
where $\hat C$ is a positive constant depending only on $x_0^i$, $d_0^i$, $E$, $\mu_i$, $\sigma^i$, $\Lambda_i$ $\gamma_i$, $\eta_i$, $T$.

Finally, regarding the last part of the statement, it is easy to see that $\hat X^i$ coincides with $\hat X^{i,\hat P}$ of equation \eqref{X}, while $\hat Y^{1,i},\hat Z^{1,i,j},\hat U^{1,i,j}$ coincide with $\hat Y^{1,i,\hat P},\hat Z^{1,i,j,\hat P},\hat U^{1,i,j,\hat P}$ of equation \eqref{hatY1}. Finally, it is also clear that $\hat P$ is the equilibrium price process, as formula \eqref{Phat} follows directly from the equilibrium condition \eqref{Equilibrium} and the definition of $\hat q^{i,\hat P}$ in \eqref{qhat}.
\ep

\begin{Theorem}\label{Thm:P_Dynamics}
The equilibrium price process $\hat P=(\hat P_t)_{0\leq t\leq T}$ is a martingale. More precisely, the dynamics of $\hat P$ is given by
\begin{align}\label{hatP_dynamics}
\hat P_t \ &= \ \hat P_0 + \sum_{i=1}^N \int_0^t \frac{\bar\gamma}{\gamma_i}2Y_s^{2,i}\sigma_s^i\Big(\rho_i\,dW_s^0 + \sqrt{1 - \rho_i^2}\,dW_s^i\Big) + \sum_{i=0}^N \int_0^t \bigg(\sum_{j=1}^N \frac{\bar\gamma}{\gamma_j} \hat Z_s^{1,j,i}\bigg)dW_s^i \\
&\quad \ + \sum_{i=1}^N \int_{(0,t]\times E} \bigg(\frac{\bar\gamma}{\gamma_i}2(D_s^i-\hat X_s^i)U_s^{2,i}(e) + \sum_{j=1}^N \frac{\bar\gamma}{\gamma_j} \hat U_s^{1,j,i}(e)\bigg)\,(\pi^i-\nu^i)(ds,de) \notag,
\end{align}
for all $0\leq t\leq T$. Similarly, the optimal trading strategies $\hat q^{1,\hat P},\ldots,\hat q^{N,\hat P}$ are martingales.
\end{Theorem}
\textbf{Proof.}
Recall that
\[
\hat P_t \ = \  \sum_{i=1}^N \frac{\bar\gamma}{\gamma_i} \Big(2 Y_t^{2,i}\big(D_t^i - \hat X_t^i\big) + \hat Y_t^{1,i}\Big), \qquad 0\leq t\leq T,
\]
where $D^i$, $Y^{2,i}$, $\hat X^i$, $\hat Y^{1,i}$ satisfy respectively equations \eqref{Di}, \eqref{hatY2}, \eqref{CoupledFBSDEX}, \eqref{CoupledFBSDEY}. Then, an application of It\^o's formula yields
\[
\hat P_t \ = \ \hat P_0 + \int_0^t \hat b_s\,ds + \textup{martingale},
\]
with the martingale term as in \eqref{hatP_dynamics} and
\[
\hat b_t \ = \ \sum_{i=1}^N \frac{\bar\gamma}{\gamma_i} \Big(2Y_t^{2,i}\,\big(\mu_i - \hat q_t^{i,\hat P}\big) - 2f_t^{2,i}\,\big(D_t^i - \hat X_t^i\big) - f_t^{1,i,\hat P}\Big),
\]
where recall that ($\hat q^i$ stands for $\hat q^{i,\hat P}$)
\begin{align}
\hat q_t^i \ &= \ \frac{1}{2\gamma_i} \Big(2 Y_t^{2,i}\big(D_t^i - \hat X_t^i\big) + \hat Y_t^{1,i} - \hat P_t\Big),\label{hat_q}	\\
f_t^{2,i} \ &= \ -\frac{1}{\gamma_i}|Y_t^{2,i}|^2, \notag \\
f_t^{1,i,\hat P} \ &= \ 2\mu_i Y_t^{2,i} + \frac{1}{\gamma_i}Y_t^{2,i}\big(\hat P_t - \hat Y_t^{1,i}\big). \notag
\end{align}
Hence, $\hat b_t$ can be rewritten as
\begin{align*}
\hat b_t \ &= \ \sum_{i=1}^N \frac{\bar\gamma}{\gamma_i} \bigg(-2\hat q_t^iY_t^{2,i} + \frac{2}{\gamma_i}|Y_t^{2,i}|^2\,\big(D_t^i - \hat X_t^i\big) - \frac{1}{\gamma_i}Y_t^{2,i}\big(\hat P_t - \hat Y_t^{1,i}\big)\bigg) \\
&= \ \sum_{i=1}^N \frac{\bar\gamma}{\gamma_i} \bigg(-2\hat q_t^iY_t^{2,i} + \frac{1}{\gamma_i}Y_t^{2,i}\Big(2Y_t^{2,i}\,\big(D_t^i - \hat X_t^i\big) - \big(\hat P_t - \hat Y_t^{1,i}\big)\Big)\bigg).
\end{align*}
By the expression of $\hat q_t^i$ in \eqref{hat_q}, we find
\[
\hat b_t \ = \ \sum_{i=1}^N \frac{\bar\gamma}{\gamma_i} \big(-2\hat q_t^iY_t^{2,i} + 2\hat q_t^iY_t^{2,i}\big) \ = \ 0,
\]
which proves that $\hat P$ is a martingale. Finally, let us consider an optimal trading strategy $\hat q^i$. By formula \eqref{hat_q}, we have
\[
\hat q_t^i \ = \ \hat q_0^i + \int_0^t \hat b_s^i\,ds + \textup{martingale},
\]
with
\[
\hat b_t^i \ = \ \frac{1}{2\gamma_i} \Big(2Y_t^{2,i}\,\big(\mu_i - \hat q_t^i\big) - 2f_t^{2,i}\,\big(D_t^i - \hat X_t^i\big) - f_t^{1,i,\hat P}\Big),
\]
where we used the fact that $\hat P$ is a martingale. Then, we see that proceeding along the same lines as for $\hat P$ we deduce that $\hat q^i$ is a martingale.
\ep

\vspace{3mm}

We now provide a formula for the solution to the coupled forward-backward system of equations \eqref{CoupledFBSDEX}-\eqref{CoupledFBSDEY}. To this end, the following formula \eqref{hatY1_formula_jumps} for the $Y$-component turns out to be particularly useful, especially in the case without jumps (as it will be shown in the next section). In the general case, formula \eqref{hatY1_formula_jumps} provides compact expressions for both the equilibrium price and the forward process, see formulae \eqref{Phat_matrix_general} and \eqref{SystemX_general} of Proposition \ref{P:Y1_formula_jumps}. In particular, formula \eqref{Phat_matrix_general} for the equilibrium price allows in turn to find a more explicit formula for the optimal trading rates in the case $\mu_i=0$ for every $i$ (see Corollary \ref{C:General}).

\begin{Proposition}\label{P:Y1_formula_jumps}
The following formula holds (notice that $1-\bar\gamma\,\theta_t\neq0$, for every $0\leq t\leq T$):
\begin{equation}\label{hatY1_formula_jumps}
\boldsymbol{\hat Y}_t^1 \ = \ \frac{\bar\gamma}{1 - \bar\gamma\,\theta_t}\boldsymbol a_t\,\mathbf 1_N\trans\,\mathbf J\,\big(2\,\boldsymbol\Delta_t + 2\,\boldsymbol{\tilde a}_t + \bar\gamma\,\mathbf J\,\boldsymbol b_t\big) + 2\,\boldsymbol{\tilde a}_t + \bar\gamma\,\mathbf J\,\boldsymbol b_t,
\end{equation}
for all $0\leq t\leq T$, where $\bar\gamma$ is as in \eqref{Phat}, $\theta_t=a_t^1/\gamma_1+\cdots+a_t^N/\gamma_N$ with $a_t^i$ as in \eqref{ai}, $\mathbf 1_N\trans$ denotes the transpose of the column vector $\mathbf 1_N$ with all entries equal to one, while $\boldsymbol{\hat Y}_t^1$, $\boldsymbol\Delta_t$, $\boldsymbol a_t$, $\boldsymbol{\tilde a}_t$, $\boldsymbol b_t$ are column vectors of dimension $N$ given by
\begin{equation}\label{vectors}
\boldsymbol{\hat Y}_t^1 =
\!\left(\!\begin{array}{c}
\hat Y_t^{1,1} \\
\vdots \\
\hat Y_t^{1,N}
\end{array}\!\right)\!\!,
\quad
\boldsymbol\Delta_t =
\!\left(\!\begin{array}{c}
Y_t^{2,1}\big(D_t^1 - \hat X_t^1\big) \\
\vdots \\
Y_t^{2,N}\big(D_t^N - \hat X_t^N\big)
\end{array}\!\right)\!\!,
\quad
\boldsymbol a_t =
\!\left(\!\begin{array}{c}
 a_t^1 \\
\vdots \\
 a_t^N
\end{array}\!\right)\!\!,
\quad
\boldsymbol{\tilde a}_t =
\!\left(\!\begin{array}{c}
\mu_1\gamma_1 a_t^1 \\
\vdots \\
\mu_2\gamma_2 a_t^N
\end{array}\!\right)\!\!,
\quad
\boldsymbol b_t =
\!\left(\!\begin{array}{c}
 b_t^1 \\
\vdots \\
 b_t^N
\end{array}\!\right)\!\!,
\end{equation}
with
\begin{align}
 a_t^i \ &= \ \frac{1}{\gamma_i}(T-t)\,Y_t^{2,i}, \label{ai} \\
 b_t^i \ &= \ \frac{1}{\gamma_i\Gamma_t^i}\int_t^T \bigg(\int_t^s \E\big[\Gamma_r^i\kappa_r^i\big|\Fc_t\big]\,dr\bigg)ds \ = \ \frac{1}{\gamma_i\Gamma_t^i}\E\bigg[\int_t^T \Gamma_r^i\kappa_r^i\,(T-r)\,dr\bigg|\Fc_t\bigg], \label{bi} \\
\kappa_t^i \ &= \ \sum_{e\in E} U_t^{2,i}(e)\bigg(2(D_t^i-\hat X_t^i)U_t^{2,i}(e) + \sum_{j=1}^N \frac{\gamma_i}{\gamma_j}\hat U_t^{1,j,i}(e)\bigg) \lambda_i(\beta_{t-}^i,e)\,1_{\{\beta_{t-}^i\neq e\}}. \label{kappai}
\end{align}
Moreover, the $N\times N$ matrices $\mathbf A_t$ and $\mathbf J$ are defined as
\begin{equation}\label{MatrixAt}
\mathbf A_t =
\left(\begin{array}{cccc}
\boldsymbol a_t & \boldsymbol a_t & \cdots & \boldsymbol a_t
\end{array}\right) =
\left(\begin{array}{ccccc}
 a_t^1 &  a_t^1 &  a_t^1 & \cdots &  a_t^1 \\
 a_t^2 &  a_t^2 &  a_t^2 & \cdots &  a_t^2 \\
\vdots & \vdots & \vdots & \ddots & \vdots \\
 a_t^N &  a_t^N &  a_t^N & \cdots &  a_t^N \\
\end{array}\right), \qquad
\mathbf J =
\left(\begin{array}{ccccc}
 \frac{1}{\gamma_1} &  0 &  0 & \cdots &  0 \\
 0 &  \frac{1}{\gamma_2} &  0 & \cdots &  0 \\
\vdots & \vdots & \vdots & \ddots & \vdots \\
 0 &  0 &  0 & \cdots &  \frac{1}{\gamma_N} \\
\end{array}\right).
\end{equation}
In addition, the equilibrium price is given by
\begin{equation}\label{Phat_matrix_general}
\hat P_t \ = \ \frac{\bar\gamma}{1 - \bar\gamma\,\theta_t}\mathbf 1_N\trans\,\mathbf J\,\big(2\,\boldsymbol\Delta_t + 2\,\boldsymbol{\tilde a}_t + \bar\gamma\,\mathbf J\,\boldsymbol b_t\big), \qquad \text{for all }0\leq t\leq T.
\end{equation}
Finally, \eqref{CoupledFBSDEX} can be rewritten as follows:
\begin{equation}\label{SystemX_general}
d\boldsymbol{\hat X}_t \ = \ \frac{1}{2}\,\mathbf J \bigg(\mathbf I - \frac{\bar\gamma}{1 - \bar\gamma\,\theta_t}\big(\mathbf 1_{N\times N} - \mathbf A_t\big)\,\mathbf J\bigg)\big(2\,\boldsymbol\Delta_t + 2\,\boldsymbol{\tilde a}_t + \bar\gamma\,\mathbf J\,\boldsymbol b_t\big)\,dt, \qquad 0\leq t\leq T,
\end{equation}
with $\boldsymbol{\hat X}_0=\boldsymbol x_0$, where $\mathbf 1_{N\times N}$ denotes the $N\times N$ matrix with all entries equal to $1$ and
\[
\boldsymbol{\hat X}_t =
\left(\begin{array}{c}
\hat X_t^1 \\
\vdots \\
\hat X_t^N
\end{array}\right),
\qquad\qquad
\boldsymbol x_0 =
\left(\begin{array}{c}
x_0^1 \\
\vdots \\
x_0^N
\end{array}\right).
\]

\end{Proposition}
\textbf{Proof.}
We split the proof into three steps.

\vspace{1mm}

\noindent\textbf{\emph{Proof of formula \eqref{hatY1_formula_jumps}.}} We begin recalling from \eqref{yi1} that $\hat Y^{1,i}$ is given by the following formula:
\begin{align}\label{yi1_bis}
\hat Y_t^{1,i} \ &= \ \frac{1}{\Gamma_t^i}\E\bigg[\int_t^T \Gamma_s^i Y_s^{2,i}\Big(2\mu_i+\frac{1}{\gamma_i}\hat P_s\Big)ds\bigg|\Fc_t\bigg] \notag \\
&= \ \frac{1}{\Gamma_t^i}\int_t^T \bigg(2\mu_i\E\big[\Gamma_s^i Y_s^{2,i}\big|\Fc_t\big] + \frac{1}{\gamma_i}\E\big[\Gamma_s^i Y_s^{2,i} \hat P_s\big|\Fc_t\big]\bigg)ds,
\end{align}
for all $0\leq t\leq T$, where $\Gamma_t^i=e^{-\frac{1}{\gamma_i}\int_0^t  Y_s^{2,i}ds}$. Now, an application of It\^o's formula yields that the process $\Gamma^i Y^{2,i}$ is a martingale and, in particular, it holds that
\[
\Gamma_s^i Y_s^{2,i} \ = \ \Gamma_t^i Y_t^{2,i} + \int_{(t,s]\times E} \Gamma_r^i U_r^{2,i}(e)\,(\pi^i-\nu^i)(dr,de), \qquad 0\leq t\leq s\leq T.
\]
As a consequence, recalling that the dynamics of $\hat P$ is given by \eqref{hatP_dynamics}, we see that
\begin{align}\label{GammaYhatP}
&\E\big[\Gamma_s^i Y_s^{2,i} \hat P_s\big|\Fc_t\big] \notag \\
&= \ \Gamma_t^i Y_t^{2,i} \hat P_t + \E\bigg[\int_{(t,s]\times E} \Gamma_r^i U_r^{2,i}(e)\bigg(\frac{\bar\gamma}{\gamma_i}2(D_r^i-\hat X_r^i)U_r^{2,i}(e) + \sum_{j=1}^N \frac{\bar\gamma}{\gamma_j}\hat U_r^{1,j,i}(e)\bigg)\nu^i(dr,de)\bigg|\Fc_t\bigg] \notag \\
&= \ \Gamma_t^i Y_t^{2,i} \hat P_t + \frac{\bar\gamma}{\gamma_i} \E\bigg[\int_t^s \Gamma_r^i\kappa_r^i\,dr\bigg|\Fc_t\bigg],
\end{align}
where $\kappa^i$ is given by \eqref{kappai}. Hence, by the martingale property of $\Gamma^i Y^{2,i}$ and \eqref{GammaYhatP}, we can rewrite formula \eqref{yi1_bis} as follows
\begin{align}\label{eq:Y1Phat}
\hat Y_t^{1,i} \ = \ \frac{1}{\Gamma_t^i}\int_t^T \bigg(2\mu_i\Gamma_t^i Y_t^{2,i} + \frac{1}{\gamma_i}\Gamma_t^i Y_t^{2,i} \hat P_t + \frac{\bar\gamma}{\gamma_i^2} \int_t^s \E\big[\Gamma_r^i\kappa_r^i\big|\Fc_t\big]\,dr\bigg)ds \ = \  a_t^i\,\hat P_t + 2\mu_i\gamma_i\, a_t^i + \frac{\bar\gamma}{\gamma_i}\,b_t^i,
\end{align}
where $ a^i$ and $ b^i$ are given by \eqref{ai} and \eqref{bi}, respectively. Using formula \eqref{Phat} for $\hat P$, we find
\[
\hat Y_t^{1,i} \ = \ a_t^i\sum_{j=1}^N \frac{\bar\gamma}{\gamma_j} \Big(\hat Y_t^{1,j} + 2 Y_t^{2,j}\big(D_t^j - \hat X_t^j\big)\Big) + 2\mu_i\gamma_i\, a_t^i + \frac{\bar\gamma}{\gamma_i}\,b_t^i.
\]
The latter can be written in matrix form as follows
\begin{equation}\label{hatY1_Matrix}
\boldsymbol{\hat Y}_t^1 \ = \ \bar\gamma\,\mathbf A_t\,\mathbf J\,\boldsymbol{\hat Y}_t^1 + 2\,\bar\gamma\,\mathbf A_t\,\mathbf J\,\boldsymbol\Delta_t + 2\,\boldsymbol{\tilde a}_t + \bar\gamma\,\mathbf J\,\boldsymbol b_t,
\end{equation}
where $\boldsymbol\Delta_t$ is the column vector of dimension $N$ given in \eqref{vectors}. In order to solve for $\boldsymbol{\hat Y}^1$, we rewrite \eqref{hatY1_Matrix} as follows
\begin{equation}\label{hatY1_Matrix_bis}
\Big(\mathbf I - \bar\gamma\,\mathbf A_t\,\mathbf J\Big)\boldsymbol{\hat Y}_t^1 \ = \ 2\,\bar\gamma\,\mathbf A_t\,\mathbf J\,\boldsymbol\Delta_t + 2\,\boldsymbol{\tilde a}_t + \bar\gamma\,\mathbf J\,\boldsymbol b_t,
\end{equation}
where $\mathbf I$ is the $N\times N$ identity matrix. Hence, we can solve for $\boldsymbol{\hat Y}^1$ if the matrix on the left-hand side of \eqref{hatY1_Matrix_bis} is invertible. We now prove that this holds true and the inverse matrix of $\mathbf I-\bar\gamma\,\mathbf A_t\,\mathbf J$ is given by
\begin{equation}\label{InverseMatrix}
\bigg(\mathbf I - \bar\gamma\,\mathbf A_t\,\mathbf J\bigg)^{-1} \ = \ \mathbf I + \frac{\bar\gamma}{1 - \bar\gamma\,\theta_t}\mathbf A_t\,\mathbf J, \qquad \text{with }\theta_t \ = \ \sum_{i=1}^N \frac{a_t^i}{\gamma_i}.
\end{equation}
Let us first check that $1-\bar\gamma\,\theta_t\neq0$, so that \eqref{InverseMatrix} is well-defined. To this regard, notice that the $i$-th element $ a_t^i$, which is given by formula \eqref{ai}, can also be written as follows
\begin{equation}\label{ai_bis}
 a_t^i \ = \ 1 - \E\bigg[e^{-\frac{1}{\gamma_i}\int_t^T  Y_s^{2,i}ds}\bigg|\Fc_t\bigg].
\end{equation}
Let us prove equality \eqref{ai_bis}. By the definition of $\Gamma^i$, we have
\[
\Gamma_T^i \ = \ \Gamma_t^i - \frac{1}{\gamma_i}\int_t^T \Gamma_s^i Y_s^{2,i}\,ds.
\]
Taking the conditional expectation with respect to $\Fc_t$, we find
\[
\E\bigg[e^{-\frac{1}{\gamma_i}\int_t^T  Y_s^{2,i}ds}\bigg|\Fc_t\bigg] \ = \ \frac{1}{\Gamma_t^i}\E\big[\Gamma_T^i\big|\Fc_t\big] \ = \ 1 - \frac{1}{\gamma_i\Gamma_t^i}\int_t^T \E\big[\Gamma_s^i Y_s^{2,i}\big|\Fc_t\big]\,ds.
\]
By the martingale property of $\Gamma^i Y^{2,i}$, we see that
\begin{equation}\label{1-a}
\E\bigg[e^{-\frac{1}{\gamma_i}\int_t^T  Y_s^{2,i}ds}\bigg|\Fc_t\bigg] \ = \ 1 -  a_t^i,
\end{equation}
from which \eqref{ai_bis} follows. Now, multiplying the above equality by $1/\gamma_i$ and summing with respect to $i$, we obtain
\[
\frac{1}{\bar\gamma} - \theta_t \ = \ \sum_{i=1}^N \E\bigg[e^{-\frac{1}{\gamma_i}\int_t^T  Y_s^{2,i}ds}\bigg|\Fc_t\bigg],
\]
namely $1-\bar\gamma\theta_t=\bar\gamma\sum_{i=1}^N \E[\exp(-\frac{1}{\gamma_i}\int_t^T  Y_s^{2,i}ds)|\Fc_t]$. Recalling from Proposition \ref{P:BSDE1} that $Y^{2,i}$ is non-negative and belongs to $\mathbf S^\infty(0,T)$, we deduce that $1-\bar\gamma\theta_t$ is a strictly positive real number. This shows that \eqref{InverseMatrix} is well-defined.

Let us now prove that the matrix on the right-hand side of \eqref{InverseMatrix} is the inverse matrix of $\mathbf I-\bar\gamma\,\mathbf A_t\,\mathbf J$. To this end, notice that
\begin{equation}\label{eigenvectorAJ}
(\mathbf A_t\,\mathbf J)^2 \ = \ \theta_t\,\mathbf A_t\,\mathbf J,
\end{equation}
where we recall that $\theta_t=a_t^1/\gamma_1+\cdots+a_t^N/\gamma_N$, namely $\theta_t$ is the trace of the matrix $\mathbf A_t\,\mathbf J$. Then, by direct calculation, it is easy to see that
\[
\bigg(\mathbf I + \frac{\bar\gamma}{1 - \bar\gamma\,\theta_t}\mathbf A_t\,\mathbf J\bigg)\bigg(\mathbf I - \bar\gamma\,\mathbf A_t\,\mathbf J\bigg) \ = \ \mathbf I,
\]
which shows the validity of \eqref{InverseMatrix}. This allows us to solve for $\boldsymbol{\hat Y}^1$ in \eqref{hatY1_Matrix_bis}, so that we obtain
\[
\boldsymbol{\hat Y}_t^1 \ = \ \bigg(\mathbf I + \frac{\bar\gamma}{1 - \bar\gamma\,\theta_t}\mathbf A_t\,\mathbf J\bigg)\big(2\,\bar\gamma\,\mathbf A_t\,\mathbf J\,\boldsymbol\Delta_t + 2\,\boldsymbol{\tilde a}_t + \bar\gamma\,\mathbf J\,\boldsymbol b_t\big).
\]
By \eqref{eigenvectorAJ} and the property $\mathbf A_t\,\mathbf J\,\boldsymbol v=(\sum_{i=1}^N v_i/\gamma_i)\boldsymbol a_t$, valid for every $\boldsymbol v\in\R^N$, we find
\[
\boldsymbol{\hat Y}_t^1 \ = \ 2\,\bar\gamma\bigg(1 + \frac{\bar\gamma\,\theta_t}{1 - \bar\gamma\,\theta_t}\bigg)\mathbf A_t\,\mathbf J\,\boldsymbol\Delta_t + 2\,\boldsymbol{\tilde a}_t + \bar\gamma\,\mathbf J\,\boldsymbol b_t + \frac{\bar\gamma}{1 - \bar\gamma\,\theta_t}\sum_{i=1}^N \frac{1}{\gamma_i}\left(2\mu_i\gamma_i a_t^i+\frac{\bar\gamma}{\gamma_i}b_t^i\right)\boldsymbol a_t.
\]
Since $1+\frac{\bar\gamma\,\theta_t}{1 - \bar\gamma\,\theta_t}=\frac{1}{1 - \bar\gamma\,\theta_t}$ and $\mathbf A_t\,\mathbf J\,\boldsymbol\Delta_t=\boldsymbol a_t \sum_{i=1}^N Y_t^{2,i}(D_t^i-\hat X_t^i)/\gamma_i$, this yields
\[
\boldsymbol{\hat Y}_t^1 \ = \ \frac{\bar\gamma}{1 - \bar\gamma\,\theta_t}\boldsymbol a_t\,\sum_{i=1}^N\frac{1}{\gamma_i}\bigg(2\,Y_t^{2,i}(D_t^i-\hat X_t^i) + 2\,\mu_i\gamma_i a_t^i+\frac{\bar\gamma}{\gamma_i}b_t^i\bigg) + 2\,\boldsymbol{\tilde a}_t + \bar\gamma\,\mathbf J\,\boldsymbol b_t.
\]
Finally, noting that $\sum_{i=1}^N\frac{1}{\gamma_i}(2Y_t^{2,i}(D_t^i-\hat X_t^i) + 2\mu_i\gamma_i a_t^i+\bar\gamma b_t^i/\gamma_i)=\mathbf 1_N\trans\,\mathbf J\,(2\,\boldsymbol\Delta_t + 2\,\boldsymbol{\tilde a}_t + \bar\gamma\,\mathbf J\,\boldsymbol b_t)$, we conclude that formula \eqref{hatY1_formula_jumps} holds.

\vspace{1mm}

\noindent\textbf{\emph{Proof of formula \eqref{Phat_matrix_general}.}} Rewriting \eqref{Phat} in matrix form, we obtain
\[
\hat P_t \ = \ \bar\gamma\,\mathbf 1_N\trans\,\mathbf J\,\Big(2 \boldsymbol\Delta_t + \boldsymbol{\hat Y}_t^1\Big), \qquad \text{for all }0\leq t\leq T.
\]
Plugging formula \eqref{hatY1_formula_jumps} into the above equality, we find
\[
\hat P_t \ = \ \bar\gamma\,\mathbf 1_N\trans\,\mathbf J\,\bigg(2\,\boldsymbol\Delta_t + \frac{\bar\gamma}{1 - \bar\gamma\,\theta_t}\boldsymbol a_t\,\mathbf 1_N\trans\,\mathbf J\,\bigg(2\,\boldsymbol\Delta_t + 2\,\boldsymbol{\tilde a}_t + \bar\gamma\,\mathbf J\,\boldsymbol b_t\bigg) + 2\,\boldsymbol{\tilde a}_t + \bar\gamma\,\mathbf J\,\boldsymbol b_t\bigg).
\]
Notice that $\mathbf 1_N\trans\,\mathbf J\,\boldsymbol a_t=\theta_t$, so that
\begin{align*}
\hat P_t \ &= \ \bar\gamma\,\frac{\bar\gamma\,\theta_t}{1 - \bar\gamma\,\theta_t}\mathbf 1_N\trans\,\mathbf J\,\big(2\,\boldsymbol\Delta_t + 2\,\boldsymbol{\tilde a}_t + \bar\gamma\,\mathbf J\,\boldsymbol b_t\big) + \bar\gamma\,\mathbf 1_N\trans\,\mathbf J\,\big(2\,\boldsymbol\Delta_t + 2\,\boldsymbol{\tilde a}_t + \bar\gamma\,\mathbf J\,\boldsymbol b_t\big) \\
&= \ \frac{\bar\gamma}{1 - \bar\gamma\,\theta_t}\mathbf 1_N\trans\,\mathbf J\,\big(2\,\boldsymbol\Delta_t + 2\,\boldsymbol{\tilde a}_t + \bar\gamma\,\mathbf J\,\boldsymbol b_t\big),
\end{align*}
which proves formula \eqref{Phat_matrix_general}.

\vspace{1mm}

\noindent\textbf{\emph{Proof of formula \eqref{SystemX_general}.}} We recall from Theorem \ref{Thm:Main2} that $\hat X^i$ solves the following ordinary differential equation with stochastic coefficients:
\[
d\hat X_t^i \ = \ \frac{1}{2\gamma_i} \big(2 Y_t^{2,i}\big(D_t^i - \hat X_t^i\big) + \hat Y_t^{1,i} - \hat P_t\big)\,dt, \qquad 0\leq t\leq T.
\]
The latter can be written in matrix as follows:
\[
d\boldsymbol{\hat X}_t \ = \ \frac{1}{2}\,\mathbf J\,\big(2\,\boldsymbol\Delta_t + \boldsymbol{\hat Y}_t^1 - \mathbf 1_N\,\hat P_t\big)\,dt.
\]
Using the expressions of $\boldsymbol{\hat Y}_t^1$ and $\hat P_t$ in \eqref{hatY1_formula_jumps} and \eqref{Phat_matrix_general}, respectively, we find
\[
d\boldsymbol{\hat X}_t \ = \ \frac{1}{2}\,\mathbf J \bigg(2\,\boldsymbol\Delta_t + 2\,\boldsymbol{\tilde a}_t + \bar\gamma\,\mathbf J\,\boldsymbol b_t - \frac{\bar\gamma}{1 - \bar\gamma\,\theta_t}\big(\mathbf 1_N\,\mathbf 1_N\trans - \boldsymbol a_t\,\mathbf 1_N\trans\big)\,\mathbf J\,\big(2\,\boldsymbol\Delta_t + 2\,\boldsymbol{\tilde a}_t + \bar\gamma\,\mathbf J\,\boldsymbol b_t\big)\bigg).
\]
Noting that $\boldsymbol a_t\,\mathbf 1_N\trans=\mathbf A_t$ and $\mathbf 1_N\,\mathbf 1_N\trans=\mathbf 1_{N\times N}$ (where we recall that $\mathbf 1_{N\times N}$ denotes the $N\times N$ matrix with all entries equal to $1$), we obtain
\begin{align*}
d\boldsymbol{\hat X}_t \ &= \ \frac{1}{2}\,\mathbf J \bigg(2\,\boldsymbol\Delta_t + 2\,\boldsymbol{\tilde a}_t + \bar\gamma\,\mathbf J\,\boldsymbol b_t - \frac{\bar\gamma}{1 - \bar\gamma\,\theta_t}\big(\mathbf 1_{N\times N} - \mathbf A_t\big)\,\mathbf J\,\big(2\,\boldsymbol\Delta_t + 2\,\boldsymbol{\tilde a}_t + \bar\gamma\,\mathbf J\,\boldsymbol b_t\big)\bigg) \\
&= \ \frac{1}{2}\,\mathbf J \bigg(\mathbf I - \frac{\bar\gamma}{1 - \bar\gamma\,\theta_t}\big(\mathbf 1_{N\times N} - \mathbf A_t\big)\,\mathbf J\bigg)\big(2\,\boldsymbol\Delta_t + 2\,\boldsymbol{\tilde a}_t + \bar\gamma\,\mathbf J\,\boldsymbol b_t\big),
\end{align*}
which corresponds to formula \eqref{SystemX_general}.
\ep

\vspace{3mm}

Using Proposition~\ref{P:Y1_formula_jumps}, it is possible to provide more precise results for the optimal trading rates  and the equilibrium price, when agents have no systematic bias on their forecasts ($\mu_i=0$).

\begin{Corollary}\label{C:General}
Suppose that $\mu_i=0$ for every $i$.
\begin{enumerate}[\upshape (i)]
\item The optimal trading rate of agent $i$ is given by (we denote $\hat q^i:=\hat q^{i,\hat P}$)
\begin{equation}\label{eq:cor-q_general}
\hat q^i_t \ = \ \frac{1 - a_t^i}{2\gamma_i} \bigg(\frac{2\,Y_t^{2,i}\,(D_t^i - \hat X_t^i) + \frac{\bar\gamma}{\gamma_i}b_t^i}{1 - a_t^i} - \hat P_t\bigg),
\end{equation}
where $1 - a_t^i=1-\frac{1}{\gamma_i}(T-t)Y_t^{2,i}$ is strictly positive, for every $i$, as it follows from equality \eqref{1-a}, moreover $b_t^i$ is given by \eqref{bi}.
\item The equilibrium price is given by
 \begin{equation}\label{eq:cor-p_general} 
\hat P_t = \sum_{i=1}^N \pi_t^i\,\frac{2\,Y_t^{2,i}\,(D_t^i - \hat X_t^i) + \frac{\bar\gamma}{\gamma_i}b_t^i}{1 - a_t^i},
\quad
\text{where }\pi_t^i := \frac{\frac{1}{\gamma_i}(1 - a_t^i)}{\sum_{j=1}^N\frac{1}{\gamma_j}(1 - a_t^j)}.
\end{equation}
\end{enumerate}
\end{Corollary}
\textbf{Proof.} (i) When $\mu_i=0$ for every $i$, the expression of $\boldsymbol{\hat Y}^1_t$ in~\eqref{hatY1_formula_jumps} reads (notice that the vector $\boldsymbol{\tilde a}_t$ in~\eqref{hatY1_formula_jumps} is equal to zero)
\[
\boldsymbol{\hat Y}_t^1 \ = \ \frac{\bar\gamma}{1 - \bar\gamma\,\theta_t}\boldsymbol a_t\,\mathbf 1_N\trans\,\mathbf J\,\big(2\,\boldsymbol\Delta_t + \bar\gamma\,\mathbf J\,\boldsymbol b_t\big) + \bar\gamma\,\mathbf J\,\boldsymbol b_t.
\]
Similarly, the expression of $\hat P_t$ in~\eqref{Phat_matrix_general} becomes
\[
\hat P_t \ = \ \frac{\bar\gamma}{1 - \bar\gamma\,\theta_t}\mathbf 1_N\trans\,\mathbf J\,\big(2\,\boldsymbol\Delta_t + \bar\gamma\,\mathbf J^2\,\boldsymbol b_t\big).
\]
Then, it holds that
\begin{equation}\label{Y=aP}
\boldsymbol{\hat Y}_t^1 \ = \ \boldsymbol a_t\,\hat P_t + \bar\gamma\,\mathbf J\,\boldsymbol b_t.
\end{equation}
Thus, by \eqref{Yi2_formula}, \eqref{qhat}, \eqref{Y=aP}, the optimal trading rate at equilibrium can be written as (we denote $\hat q^i:=\hat q^{i,\hat P}$)
\[
\hat q^i_t \ = \ \frac{1}{2\gamma_i} \big(2\,\Delta_t^i + \hat Y_t^{1,i} - \hat P_t\big) \ = \ \frac{1}{2\gamma_i} \bigg(2\,\Delta_t^i + \frac{\bar\gamma}{\gamma_i}b_t^i - (1 - a_t^i)\,\hat P_t\bigg),
\]
which yields equality \eqref{eq:cor-q_general} recalling that $\Delta_t^i=Y_t^{2,i}(D_t^i-\hat X_t^i)$.

\vspace{1mm}

\noindent(ii) From the expression of $\hat P_t$ in~\eqref{Phat_matrix_general}, we obtain (recalling that $\boldsymbol{\tilde a}_t=0$)
\begin{align*}
\hat P_t \ = \ \frac{\bar\gamma}{1 - \bar\gamma\,\theta_t}\mathbf 1_N\trans\,\mathbf J\,\big(2\,\boldsymbol\Delta_t + \bar\gamma\,\mathbf J\,\boldsymbol b_t\big) \ &= \ \frac{1}{\sum_{j=1}^N\frac{1}{\gamma_j}(1 - a_t^j)}\,\mathbf 1_N\trans\,\mathbf J\,\big(2\,\boldsymbol\Delta_t + \bar\gamma\,\mathbf J\,\boldsymbol b_t\big) \\
&= \ \frac{1}{\sum_{j=1}^N\frac{1}{\gamma_j}(1 - a_t^j)}\sum_{i=1}^N\frac{1}{\gamma_i}\bigg(2\,Y_t^{2,i}\,(D_t^i - \hat X_t^i) + \frac{\bar\gamma}{\gamma_i}b_t^i\bigg) \\
&= \ \sum_{i=1}^N \pi_t^i\,\frac{2\,Y_t^{2,i}\,(D_t^i - \hat X_t^i) + \frac{\bar\gamma}{\gamma_i}b_t^i}{1 - a_t^i},
\end{align*}
with $\pi_t^i$ as in \eqref{eq:cor-p_general}.
\ep

\section{The case without jumps}
\label{S:NoJump}

In the present section we focus on the case where there are no jumps, so that the terminal condition of $Y^{2,i}$ is deterministic and given by $\frac{1}{2}\frac{\eta_i\,e_i}{\eta_i + e_i}$, for some fixed $e_i\in E$. We also assume that for all $i=1,\ldots,N$,  $\mu_i = 0$, meaning that market players have unbiaised forecasts of their terminal demand. In such a framework, $Y^{2,i}$ solves the following backward equation:
\[
dY_t^{2,i} \ = \ \frac{1}{\gamma_i} |Y_t^{2,i}|^2\,dt, \qquad\qquad Y_T^{2,i} \ = \ \frac{1}{2}\,\frac{\eta_i\,e_i}{\eta_i + e_i} =: \frac12 \epsilon_i.
\]
Hence, $Y^{2,i}$ is given by
\begin{equation}\label{Yi2_formula}
Y_t^{2,i} \ = \ \frac{ Y^{2,i}_T}{1 + \frac{1}{\gamma_i} Y^{2,i}_T (T-t)} =: \frac12 \frac{\epsilon_i}{1+\frac12 \phi_i (T-t)}, \qquad 0\leq t\leq T, 
\quad \text{ with } \phi_i := \frac{\epsilon_i}{\gamma_i}.
\end{equation}

\vspace{3mm}

In the present framework we can give more precise formulae, compared to Corollary \ref{C:General}, for the optimal trading rates and the equilibrium price when $\mu_i=0$; moreover, we can provide a formula for the volatility of the equilibrium price. For sake of notations, we write $\tilde W^i_t := \rho_i W^0_t + \sqrt{1-\rho_i^2} W^i_t$.

\begin{Corollary}\label{C:No_Jumps}
Suppose that $\mu_i=0$ for every $i$. 
\begin{enumerate}[\upshape (i)]
\item The equilibrium price $\hat P_t$ is given by (recall that $c_i(\hat\xi_i)=e_i\hat\xi_i^2/2$, so in particular $c_i'(\hat \xi^i_t)=e_i\hat \xi^i_t$)
\begin{align} \label{eq:cor-p} 
 &\hat P_t  =  \sum_{i=1}^N  F_i(t) c'_i(\hat \xi^i_t), \quad \text{ with } \quad   F_i(t) := \frac{G(t)}{f_i(t)},  
 \quad G(t) := \bigg( \sum_{i=1}^N  1/f_i(t) \bigg)^{-1}, \\
&  f_i(t) := \gamma_i + \frac12 \epsilon_i (T-t),  \text{ and }  \hat \xi^i_t := \frac{\eta_i}{e_i + \eta_i} (D^i_t - \hat X^i_t).
\end{align}
and the optimal trading rate $\hat q^i_t$ is given by 
\begin{align}\label{eq:cor-q0}
\hat q^{i}_t \  =  \ \frac12 \frac{ c'_i(\hat \xi^i_t)  - \hat P_t}{f_i(t)}.
\end{align}

\item The dynamics of the equilibrium price writes 
\begin{align} \label{eq:cor-dp}
 & d\hat P_t \ = \  \sum_{i=1}^N  F_i(t) \epsilon_i \sigma_t^i  d\tilde W_t^i.
\end{align}

\item In particular, the volatility $\zeta=(\zeta_t)_{t\in[0,T]}$ of the equilibrium price process is deterministic, and satisfies
\begin{align}\label{eq:vol1}
& \zeta_t^2 \ =  \ \sum_{i=1}^N  (1-\rho^2_i)  ( \epsilon_i F_i(t)  \sigma^i_t  \big)^2 +   \Big( \sum_{i=1}^N  \rho_i   \epsilon_i  F_i(t) \sigma^i_t   \Big)^2. 
\end{align}

\item Moreover, if $\sigma_i=\sigma$, $\gamma_i=\gamma$, $\rho_i=\rho$, for every $i$, then
\begin{align}\label{explizeta} 
\zeta_t^2 \ &= \ \big(1-\rho^2\big)\,\big(\sigma^2(T-t)+\sigma_0^2\big)\,G^2(t)\,\sum_{i=1}^N \bigg(\frac{2Y^{2,i}_T }{\gamma + Y^{2,i}_T (T-t)}\bigg)^2 \\
&\quad \ + \rho^2\,\big(\sigma^2(T-t)+\sigma_0^2\big)\,G^2(t)\,\bigg(\sum_{i=1}^N \frac{2Y^{2,i}_T }{\gamma + Y^{2,i}_T (T-t)}\bigg)^2. \notag
\end{align}
If in addition all players have the same cost functions, namely  $e_i=e$, for every $i$, then the volatility of the equilibrium price is a decreasing function of time and it is given by
\begin{equation}\label{zeta_decreasing}
\zeta_t^2 \ = \ 4\,\big(1-\rho^2\big)\,\frac{1}{N}\,|Y_T^2|^2\,\big(\sigma^2(T-t) + \sigma_0^2\big) + 4\,\rho^2\,|Y_T^2|^2\,\big(\sigma^2(T-t) + \sigma_0^2\big),
\end{equation}
where $Y_T^2=\frac{1}{2}\frac{\eta\,e}{\eta+e}$, for every $i$.
\end{enumerate}
\end{Corollary}
{\bf Proof.} \emph{Item (i).} By formula \eqref{eq:cor-q_general}, we have in the case without jumps ($b^i=0$)
\[
\hat q^i_t \ = \ \frac{1}{2\gamma_i} \big(2\,Y_t^{2,i}\,(D_t^i - \hat X_t^i) - (1 - a_t^i)\hat P_t\big) \ = \ \frac{1}{2\gamma_i} \bigg(\frac{c'_i(\hat \xi^i_t)}{1+\frac12 \phi_i (T-t) } - (1-a^i_t) \hat P_t\bigg),
\]
where we used that 
\begin{align*}
2 Y^{2,i}_t (D^i_t - \hat X^i_t ) \ = \ \frac{\epsilon_i (D^i_t-\hat X^i_t)}{1+\frac12 \phi_i (T-t)} \ = \ \frac{ c'_i(\hat \xi^i_t)}{1+\frac12 \phi_i (T-t)}. 
\end{align*}
Noting $1-a^i_t = 1/(1+\frac12 \phi_i (T-t))$, we get
\[
\hat q^i_t \ = \ \frac{1}{2\gamma_i} \bigg( \frac{ c'_i(\hat \xi^i_t) - \hat P_t}{1+\frac12 \phi_i (T-t) } \bigg)
\]
Thus, summing up all the trading rates, we find
\begin{align}
\hat P_t \ = \ \sum_{i=1}^N \bigg(\sum_{k=1}^N \Big(\gamma_k \Big(1+\frac12\phi_k(T-t)\Big)\Big)^{-1} \bigg)^{-1}  \frac{1}{\gamma_i(1+\frac12 \phi_i (T-t) ) } c'_i(\hat \xi^i_t)
\end{align}
with
\begin{align}
F_i(t) \ := \ \frac{G(t)}{\gamma_i(1+\frac12 \phi_i (T-t) ) }, \qquad G(t) \ := \ \bigg( \sum_{k=1}^N \Big( \gamma_k \Big(1+\frac12\phi_k(T-t)\Big) \Big)^{-1} \bigg)^{-1}.
\end{align}
\emph{Items (ii) and (iii).} Using formula \eqref{eq:cor-p} and that the noise terms are only due to $D^1_t,\ldots,D^N_t$, we find 
 \begin{align*}
d\hat P_t & =   \sum_{i=1}^N F_i(t) \epsilon_i \sigma^i_t d\tilde W^i_t,
 \end{align*} 
which corresponds to formula \eqref{eq:cor-dp}. From such a formula we immediately get \eqref{eq:vol1}.\\
\emph{Item (iv).} Formula \eqref{explizeta} is a direct consequence of \eqref{eq:vol1}. Finally, if all cost functions are identical, namely $Y_{T}^{2,i}=Y_T^2=\frac{1}{2} \frac{\eta\,e}{\eta+e}$, for every $i$, then by \eqref{explizeta} we immediately get \eqref{zeta_decreasing}, which proves that the volatility is decreasing in $t$.
\ep

\begin{Remark}\label{R:Vol}
{\rm
\noindent \begin{enumerate}[\upshape (i)]

\item The item (i) in Corollary \ref{C:No_Jumps} shows that the equilibrium price is a convex combination of the forecasted marginal cost to produce the quantity $D^i_t - \hat X^i_t$. The quantity $D^i_t - \hat X^i_t$ is the best estimator an agent can have on the quantity she will have to produce at time $T$. The weights of the convex combination are deterministic functions of time. Further, for any agent, the optimal trading strategy is simply to compare its forecasted marginal cost $c'_i(\hat \xi^i_t)$ to the equilibrium price $\hat P_t$. If the forecasted marginal cost is higher (resp. lower)  than $\hat P_t$, she buys (resp. sell). 

\item Using~\eqref{eq:cor-p}, we can rewrite $\hat P$ as
\begin{align*}
\hat P_t = S_t -   \sum_{i=1}^N F_i(t) \epsilon_i  \hat X^i_t, \quad S_t := \sum_{i=1}^N F_i(t) \epsilon_i (D^i_t  - x^i_0)
\end{align*}

The process $S_t$ is an uncontrolled process which is the {\em fondamental price} in the Almgren and Chriss \cite{Almgren01} model of intraday trading. The factors $\epsilon_i F_i(t)$ reads as the {\em permanent market impact} of each agent. Note that if agents are identical, $\hat P_t$ reduces to its fundamental component because of the market clearing condition.

\item In the formula~\eqref{eq:vol1} of the volatility, the volatility functions of demand forecasts $\sigma^i$ are supposed to be decreasing in time, reflecting the fact that closer to maturity market players know better their demand. If all the functions $F_i$ were non-increasing, it would result that the volatility would decrease, making the Samuelson's effect not holding. But, the monotonicity of the functions $F_i$ are not obvious.   Nevertheless, since the functions $F_i$ form a convex combination, it holds that:
\begin{align*}
\sum_{i=1}^N F'_i(t) = 0, \text{ for all } t \in [0,T]. 
\end{align*}
As a consequence, it cannot hold that all the functions $F_i$ have the same monotonicity on the interval $(0,T)$. If the market players are homogeneous (same cost $e_i$, same penalisation of imbalances $\eta_i$, same market access $\gamma_i$ and same dependence to common noise $\rho_i$), all functions $F_i$ are constant equal to $1/N$ and the volatility reduces to:
\begin{align*}
\zeta^2_ t & = \frac{\epsilon^2}{N^2}  \Big( \sum_{i=1}^N (1-\rho^2) (\sigma^i_t)^2  + \rho^2  \Big( \sum_{i=1}^N  \sigma^i_t \Big)^2\Big).
\end{align*}
In this homogeneous case,  the monotonicity of the volatility function is fully determined by the monotonicity of the volatility of the demand forecasts. Apart from this limiting case, the monotonicity of $\zeta^2_t$ depends on the heterogeneity of the agents.

As a consequence, if the demand forecasts have decreasing volatility (i.e. increasing quality), heterogeneity is a necessary condition for the Samuelson's effect to hold. 

\item Under the assumptions for the validity of formula  \eqref{zeta_decreasing}, we see that $\zeta_t^2$ converges to zero as $N$ goes to infinity when $\rho=0$, while the limit is strictly positive when $\rho\neq0$. This result translates in the following remark: in a market with no production shocks, prices move because agents face a common economic factor.

\item Rewriting~ \eqref{eq:cor-p}, it  holds that
\begin{align*}
F_i(t) = \Big[ \sum_{k=1}^N \frac{\gamma_i + \frac12 \epsilon_i (T-t)}{\gamma_k + \frac12 \epsilon_k (T-t)}\Big]^{-1}
\end{align*}
Hence, when there are no market frictions, i.e. all the $\gamma_i$ are zero, all the functions $F_i$ are constant, and thus, the equilibrium price still exists.
\end{enumerate}
}
\end{Remark}

\begin{table}[hbt!]
\center
\begin{tabular}{l  c  c }
 						& Type 1 		& Type 2 \\ 
						& & \\ \hline\hline
						& & \\
Figure~\ref{fig:zeta} (Left) & $\sigma_1^2 = 20$, $\sigma_0^2 = 5$,  $\rho = 0$,  &  $\sigma_1^2 = 0$, $\sigma_0^2 = 0$, $\rho = 0$,  \\
 					 & $e = 10$, $\gamma = 0.1$ $\eta=5$. &  $e = 10$, $\gamma = 100$, $\eta=5$.  \\
						& & \\
Figure~\ref{fig:zeta} (Right) & $\sigma_1^2 = 1$, $\sigma_0^2 = 0$, $\rho = 1$,  &  $\sigma_1^2 = 0.1$, $\sigma_0^2 = 10$, $\rho = -1$,  \\
 					 & $e = 2$, $\gamma = 1$ $\eta=5$. &  $e = 1$, $\gamma = 1$ $\eta=5$.  \\
\end{tabular}
\caption{Parameters value used for the Figure~\ref{fig:zeta}.}\label{tab:params}
\end{table}
\begin{figure}[hbt!]
\center
\includegraphics[width=0.4\textwidth]{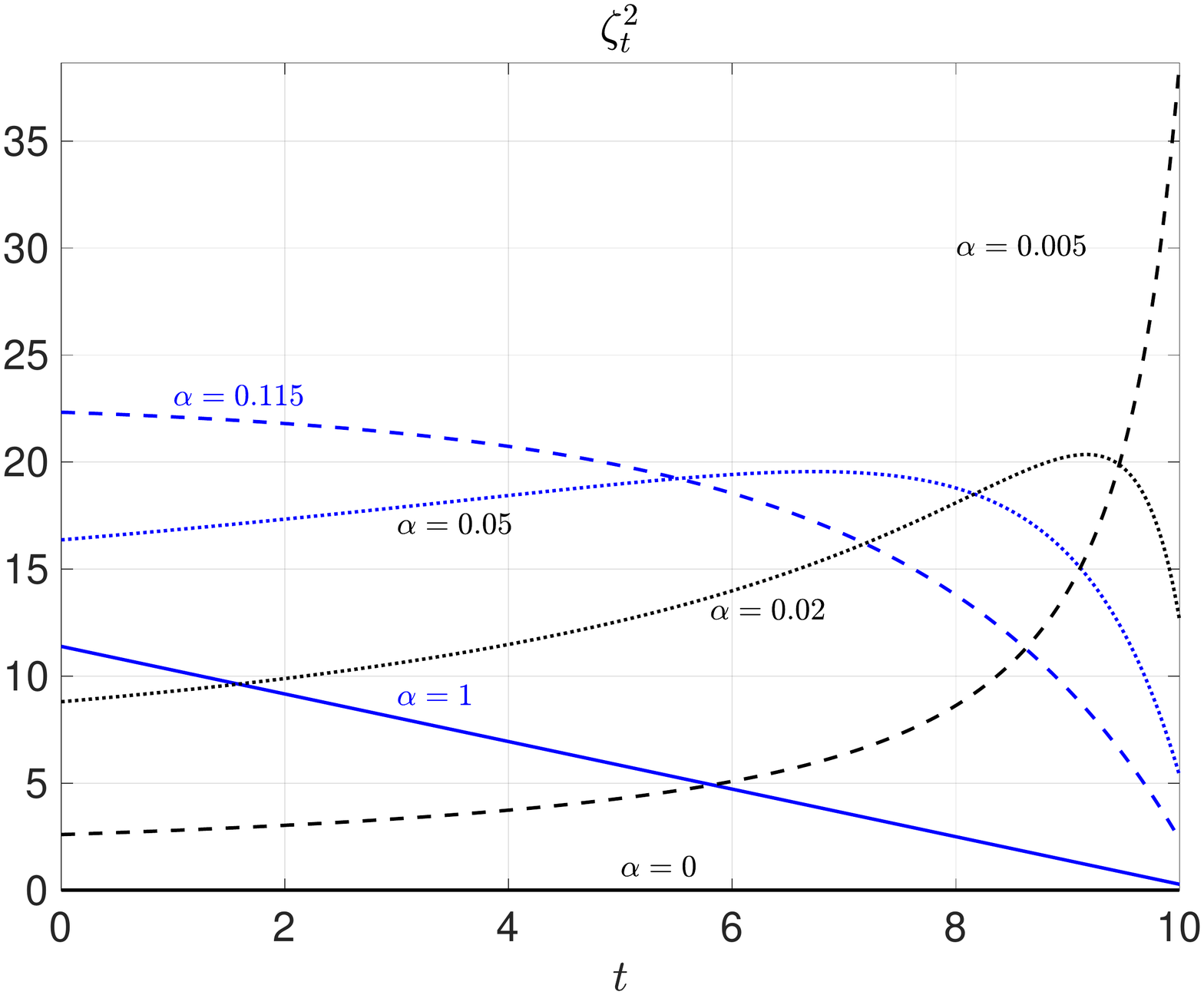}  \includegraphics[width=0.40\textwidth]{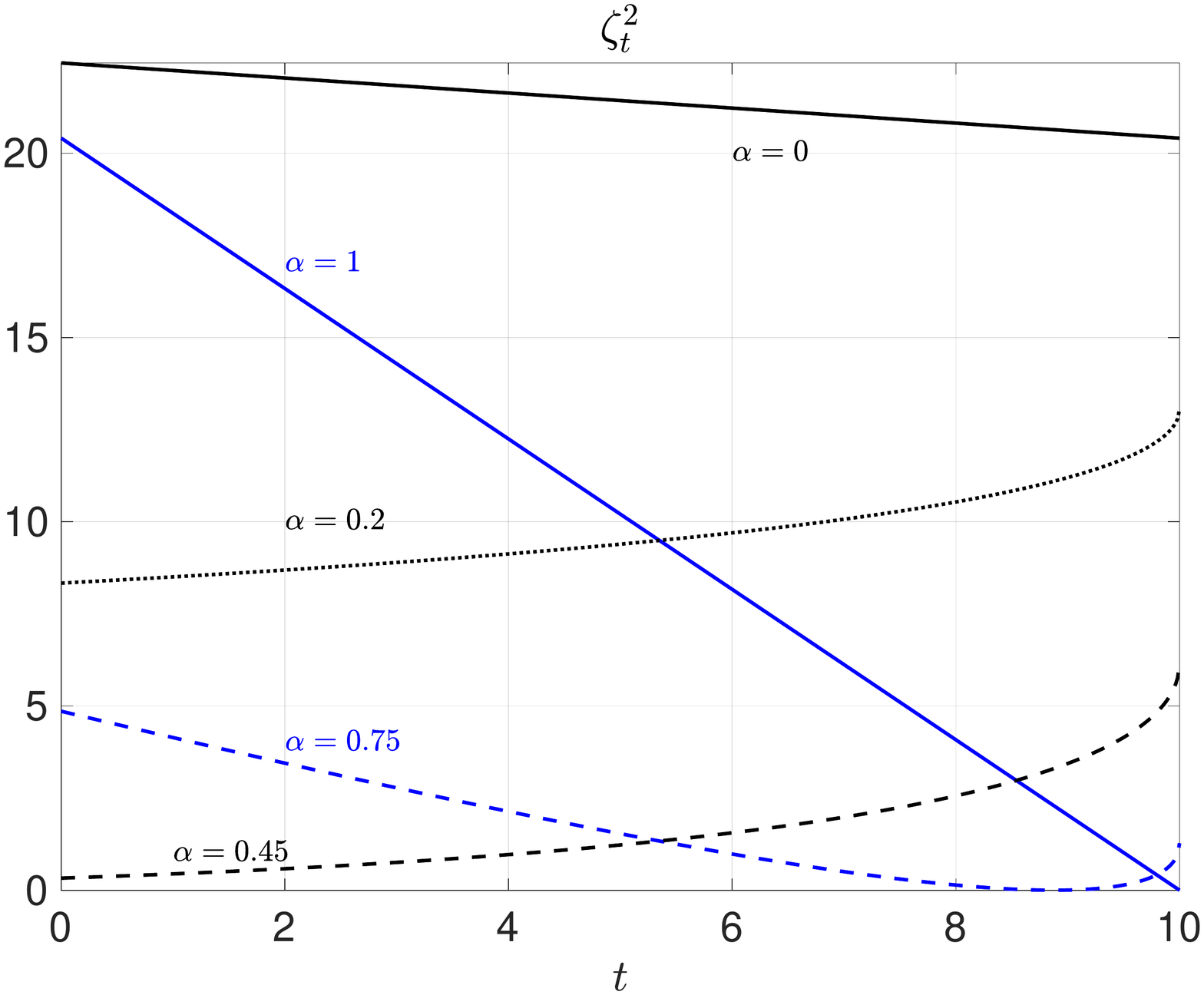}
\caption{Price volatility function $\zeta^2_t$ as a function of the proportion of type 1 agents $\alpha$ with parameters given in Table~\ref{tab:params}}
\label{fig:zeta}
\end{figure}
\paragraph{Numerical illustrations.} To illustrate how heterogeneity may induce rich behaviour of the price volatility, we consider the case of mixing two types of agents with characteristics given by Table~\ref{tab:params}.  Parameters value do not pretend to have any significant meaning compared to an observed market and are only provided for illustration of a potential behaviour. We consider two cases. In the first one, agents are not affected by a common noise while in the second case, the first type of agent is positively affected by the common noise and the second case is negatively affected.  Figure~\ref{fig:zeta} (Left) illustrates the first situation and Figure~\ref{fig:zeta} (Right) the second one. The fraction $\alpha$ designates the proportion of agents of type~1. 

In the first situation, we observe that when there are  only agents of type~1, the volatility is decreasing. As we introduce more and more agents of type 2, the volatility is still decreasing but becomes concave ($\alpha = 0.115$). Then, passed a certain threshold, the volatility is no longer monotonic and starts to increase. For a large amount of agents of type~2 ($\alpha=0.005$), the volatility is purely increasing. In the end, when there are  only agents of type~2, the volatility is zero because $\sigma^2_t \equiv 0$.

In the second situation, the same phenomenon occurs but at a higher proportion of agents of type~2. As soon as there almost half agents of type~2, the volatility becomes increasing.

\section{The case with jumps}
\label{S:Jumps}

In the present section we consider the case with jumps only and make the following assumptions.
\begin{enumerate}[a)]
\item for every $i=1,\ldots,N$, the demand forecast is perfect and $\sigma^i=0$;
\item the set $E=\{g,b\}\subset(0,+\infty)$ is made of two states ($g$ stands for \emph{good} and $b$ for \emph{bad}), with $g<b$;
\item for every $i=1,\ldots,N$, the Markov chain $\beta^i$ has state space $E=\{g,b\}$, initial state $g$ at time $t=0$ and intensity matrix given by
\[
\Lambda_i \ = \
\left(
\begin{array}{cc}
- \lambda_i & \lambda_i \\
0 & 0
\end{array}
\right),
\]
where $\lambda_i$ is a fixed strictly positive real number. In other words, each agent has an intensity rate $\lambda^i$ to jump from the good state to the bad state and a zero intensity rate to jump from the bad state to the good state (so, in particular, if an agent is in the bad state, then she/he stays in the bad state);
\end{enumerate}
In this framework, for every $i=1,\ldots,N$, the Riccati type system of equations \eqref{Riccati} becomes
\begin{align}
y_{i,g}'(t) \ &= \ \frac{1}{\gamma_i}|y_{i,g}(t)|^2 + \lambda_i y_{i,g}(t) - \lambda_i y_{i,b}(t), \qquad\qquad y_{i,g}(T) \ = \ \frac{1}{2}\,\frac{\eta_i\,g}{\eta_i + g}, \label{Riccati_g} \\
y_{i,b}'(t) \ &= \ \frac{1}{\gamma_i}|y_{i,b}(t)|^2, \hspace{5.2cm} y_{i,b}(T) \ = \ \frac{1}{2}\,\frac{\eta_i\,b}{\eta_i + b}. \label{Riccati_b}
\end{align}
We recall that the backward stochastic differential equation \eqref{hatY2}, driven by the Markov chain $\beta^i$, is such that
\begin{equation}\label{Y^2,i}
Y_t^{2,i} \ = \ y_{i,\beta_t^i}(t).
\end{equation}
In the case with jumps, \eqref{hatY1_formula_jumps} is not an explicit formula for $\boldsymbol{\hat Y}^1$, as a matter of fact the quantity $\boldsymbol b$ depends on $\sum_{j=1}^N\hat U^{1,j,i}$ (see \eqref{bi} and \eqref{kappai}) and therefore on $\boldsymbol{\hat Y}^1$ itself. However, if $\gamma_i=\gamma$ then formula \eqref{hatY1_formula_jumps} becomes
\[
\boldsymbol{\hat Y}_t^1 \ = \ \frac{1}{N - \text{tr}(\mathbf A_t)}\boldsymbol a_t\,\mathbf 1_N\trans\,\bigg(2\,\boldsymbol\Delta_t + 2\,\boldsymbol{\tilde a}_t + \frac{1}{N}\,\boldsymbol b_t\bigg) + 2\,\boldsymbol{\tilde a}_t + \frac{1}{N}\,\boldsymbol b_t.
\]
Recalling estimate \eqref{Estimate-coupled-system}, we see that $\sum_{i=1}^N\E[\int_0^T|b_t^i|^2dt]$ is bounded by a constant which is independent of $N$. As a consequence, the quantities $b_t^i/N$ and $\boldsymbol b_t/N$ appearing in formula \eqref{hatY1_formula_jumps} can be neglected for $N$ large enough, so, in particular, we obtain an approximate formula for $\boldsymbol{\hat Y}^1$ which is explicit.

\vspace{2mm}

Differently, consider the case with only two players, so $N=2$, moreover $\lambda_1=0$ and $\lambda_2>0$. Under those assumptions, we are able to determine the equilibrium price process together with the optimal trading strategies.

\begin{Proposition}\label{P:Jumps}
Suppose that assumptions a)-b)-c) stated at the beginning of this section hold true. Moreover, assume that $N=2$, $\lambda_1=0$ and $\lambda_2>0$.\\
Firstly, consider the following function of $t$ and $x$, which is linear in $x$:
\begin{align}\label{hatY11}
\ell(t,x) \ &:= \ \frac{\bar\gamma}{1-\frac{\bar\gamma}{\gamma_1}a_t^1-\frac{\bar\gamma}{\gamma_2}a_t^2} \bigg(\frac{1}{\gamma_1}\big(2\,Y_t^{2,1}(D_t^1-x) + 2\,\mu_1\gamma_1 a_t^1\big) \\
&\quad \ + \frac{1}{\gamma_2}\big(2\,Y_t^{2,2}(D_t^2-x_0^2 - x_0^1 + x) + 2\,\mu_2\gamma_2 a_t^2\big)\bigg)a_t^1 + 2\mu_1\gamma_1 a_t^1, \notag
\end{align}
with $\bar\gamma=\frac{1}{\frac{1}{\gamma_1}+\frac{1}{\gamma_2}}$, $Y^{2,i}$ given by \eqref{Y^2,i} and
\[
a_t^i \ = \ \frac{1}{\gamma_i}(T-t)Y_t^{2,i}, \qquad \text{for }i=1,2.
\]
Now, the process $\hat X^1=(\hat X_t^1)_{0\leq t\leq T}$ satisfying \eqref{CoupledFBSDEX} is the solution to the following linear ordinary differential equation:
\begin{equation}\label{ODE_X1_Prop}
d\hat X_t^1 \ = \ \frac{1}{\gamma_1} Y_t^{2,1}\big(D_t^1 - \hat X_t^1\big)\,dt + \mu_1 a_t^1\,dt + \frac{1}{2\gamma_1}\bigg(1 - \frac{1}{a_t^1}\bigg)\ell(t,\hat X_t^1)\,dt, \qquad \hat X_0^1 \ = \ x_0^1.
\end{equation}
Moreover, the equilibrium price process is given by
\begin{equation}\label{Price}
\hat P_t \ = \ \frac{1}{a_t^1}\ell(t,\hat X_t^1) - 2\mu_1\gamma_1 a_t^1.
\end{equation}
Finally, the optimal trading strategies are as follows:
\begin{equation}\label{OptTradStrat}
\hat q_t^1 \ = \ \frac{1}{2\gamma_1} \Big(2 Y_t^{2,1}\big(D_t^1 - \hat X_t^1\big) + \ell(t,\hat X_t^1) - \hat P_t\Big), \qquad\qquad \hat q^2 \, = \, - \hat q^1.
\end{equation}
\end{Proposition}
\textbf{Proof.}
Since $\lambda_1=0$ it follows that $\kappa^1$ in \eqref{kappai} is identically zero, therefore $b^1$ is also equal to zero. Moreover, as $N=2$, the equilibrium condition \eqref{Equilibrium} gives $\hat q^1=-\hat q^2$, so that
\begin{equation}\label{X2=X1}
\hat X_t^2 \ = \ x_0^2 + x_0^1 - \hat X_t^1, \qquad 0\leq t\leq T.
\end{equation}
As a consequence, by formula \eqref{hatY1_formula_jumps}, for $i=1$, we obtain
\[
\hat Y_t^{1,1} \ = \ \ell(t,\hat X_t^1),
\]
with $\ell$ as in \eqref{hatY11}. Now, by \eqref{eq:Y1Phat} we have
\begin{equation}\label{P=Y1}
\hat P_t \ = \ \frac{1}{a_t^1}\hat Y_t^{1,1} - 2\mu_1\gamma_1 a_t^1,
\end{equation}
which corresponds to equality \eqref{Price}. Moreover, recalling formula \eqref{qhat}, namely
\[
\hat q_t^1 \ = \ \frac{1}{2\gamma_1} \Big(2 Y_t^{2,1}\big(D_t^1 - \hat X_t^1\big) + \hat Y_t^{1,1} - \hat P_t\Big),
\]
we see that \eqref{OptTradStrat} holds true. It remains to prove that the process $\hat X^1$ is the solution to equation \eqref{ODE_X1_Prop}. To this end, we recall from Theorem \ref{Thm:Main2} that $\hat X^1$ solves the following equation
\[
d\hat X_t^1 \ = \ \frac{1}{2\gamma_1} \big(2 Y_t^{2,1}\big(D_t^1 - \hat X_t^1\big) + \hat Y_t^{1,1} - \hat P_t\big)\,dt.
\]
Using relation \eqref{P=Y1}, equality $\hat Y^{1,1}=\ell(t,\hat X_t^1)$ and also \eqref{X2=X1}, we see that the above equation coincides exactly with equation \eqref{ODE_X1_Prop}.
\ep

\begin{figure}[thbt!]
\center
\includegraphics[width=0.3\textwidth]{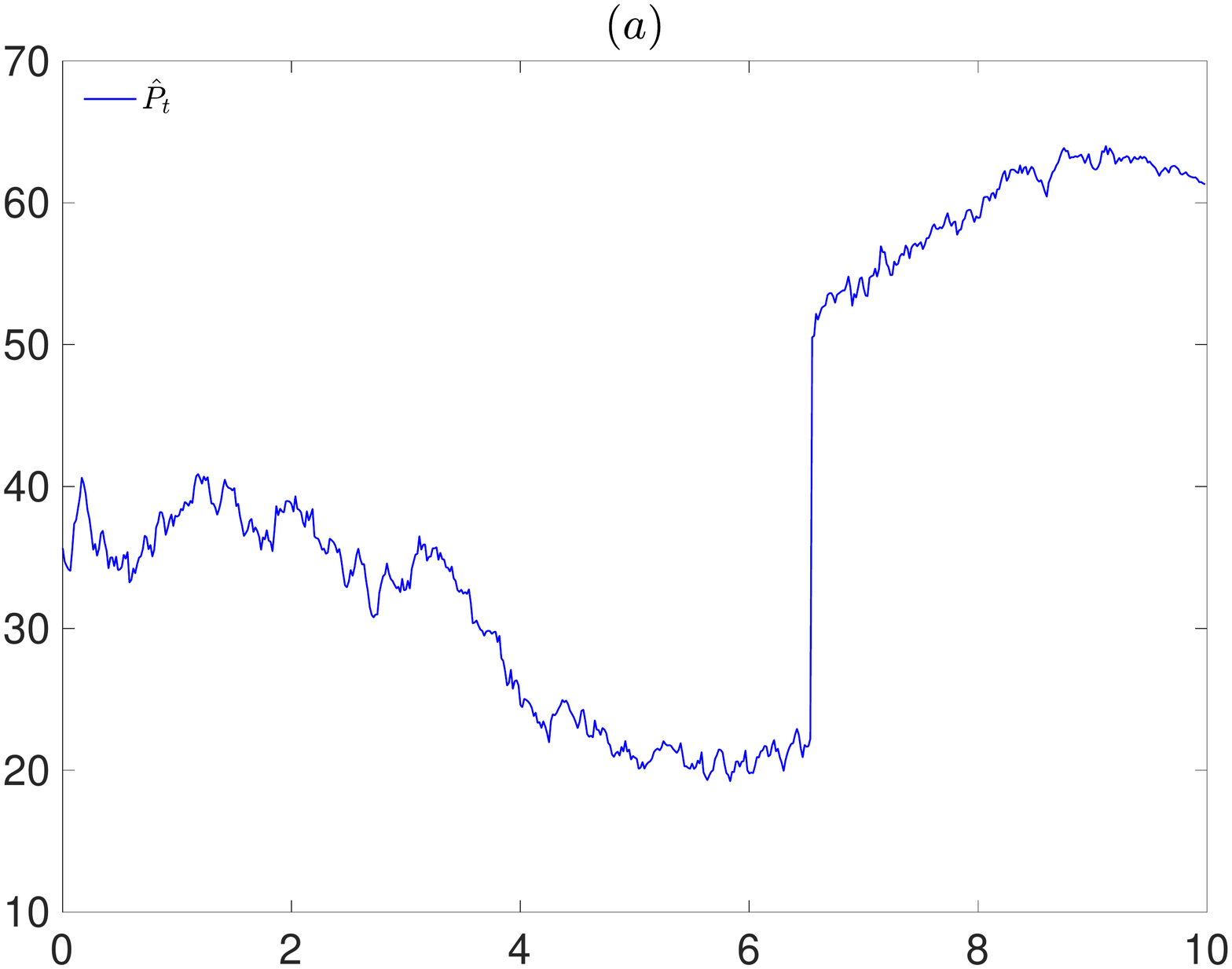}\includegraphics[width=0.3\textwidth]{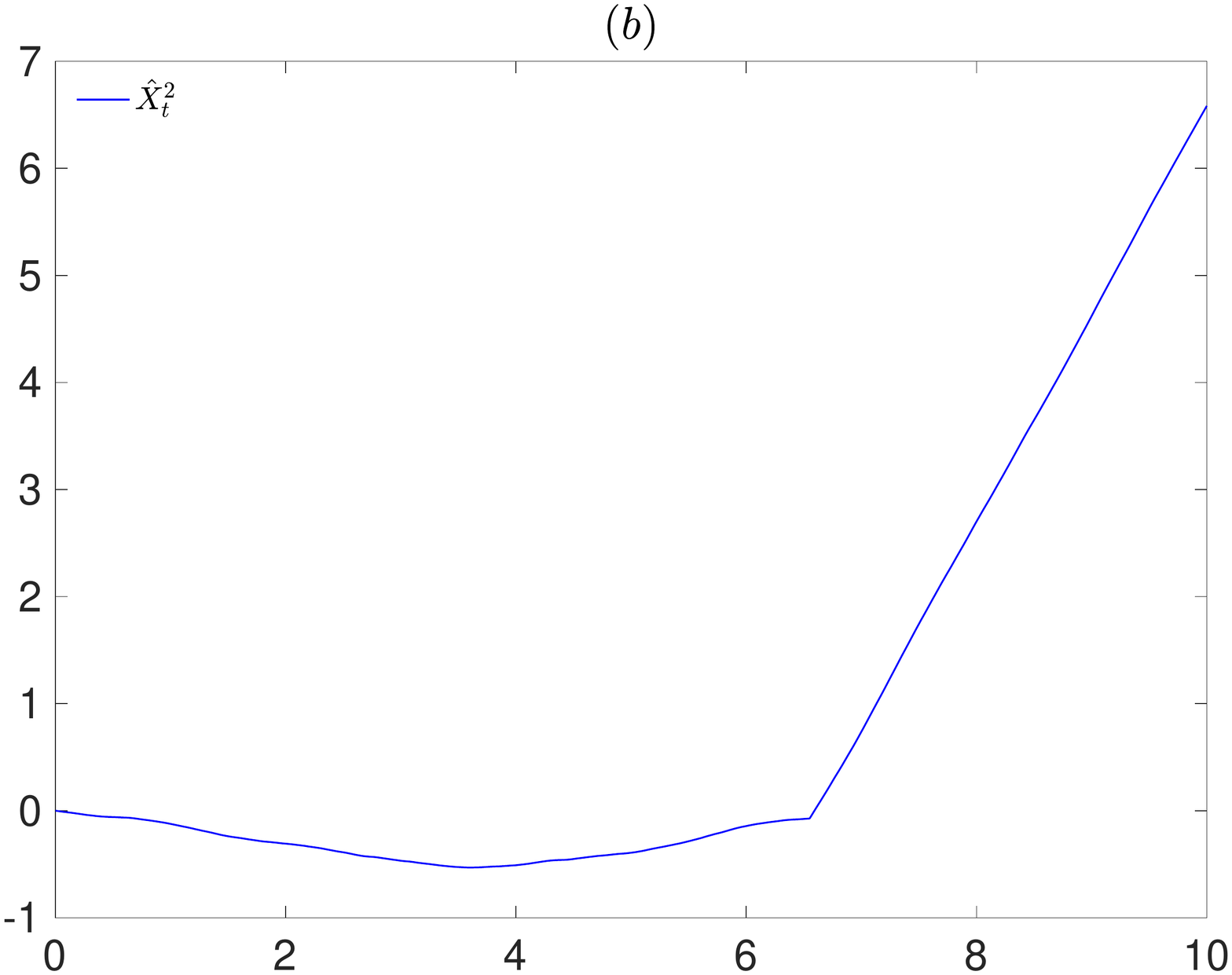} \includegraphics[width=0.3\textwidth]{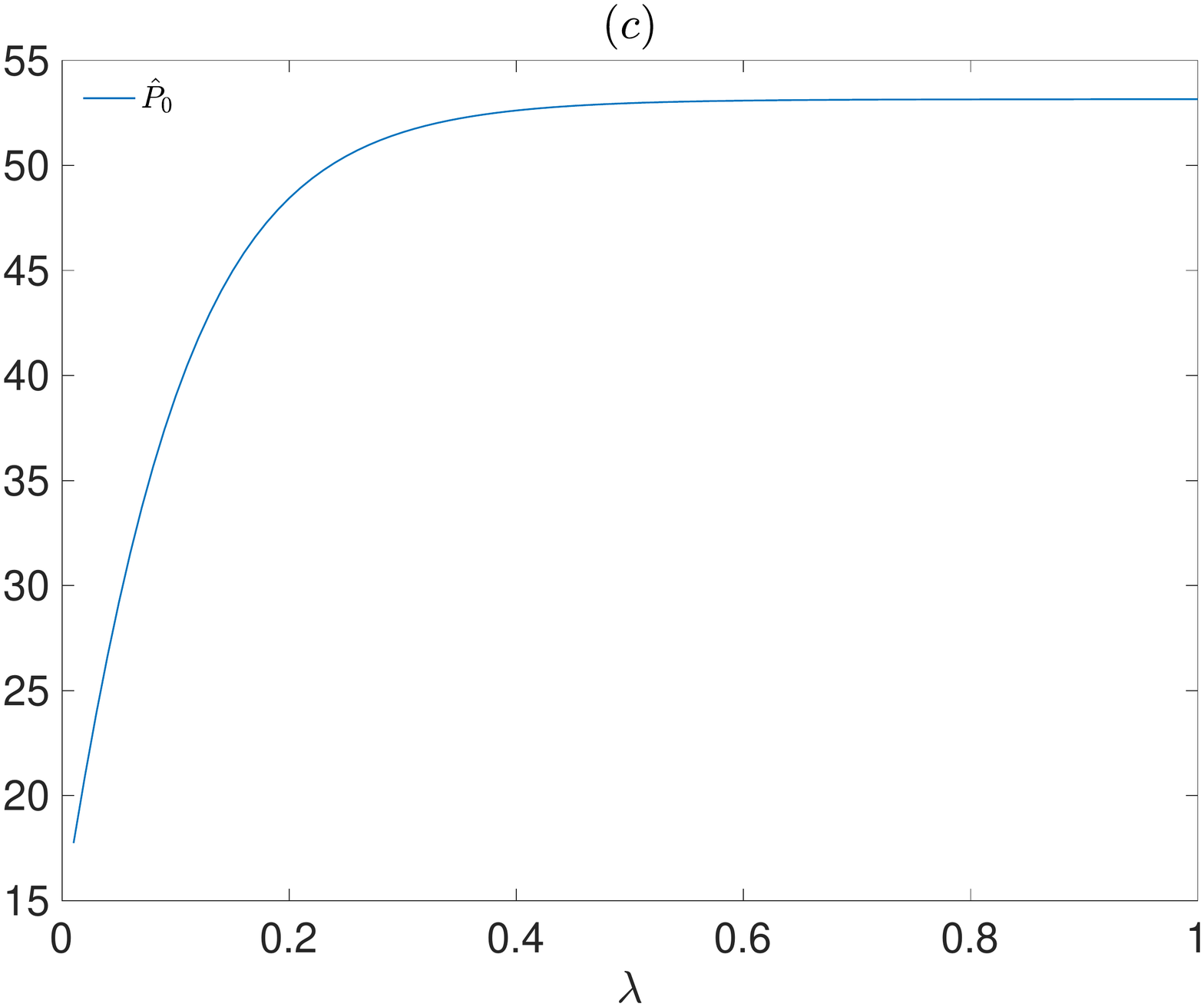}\\
\includegraphics[width=0.3\textwidth]{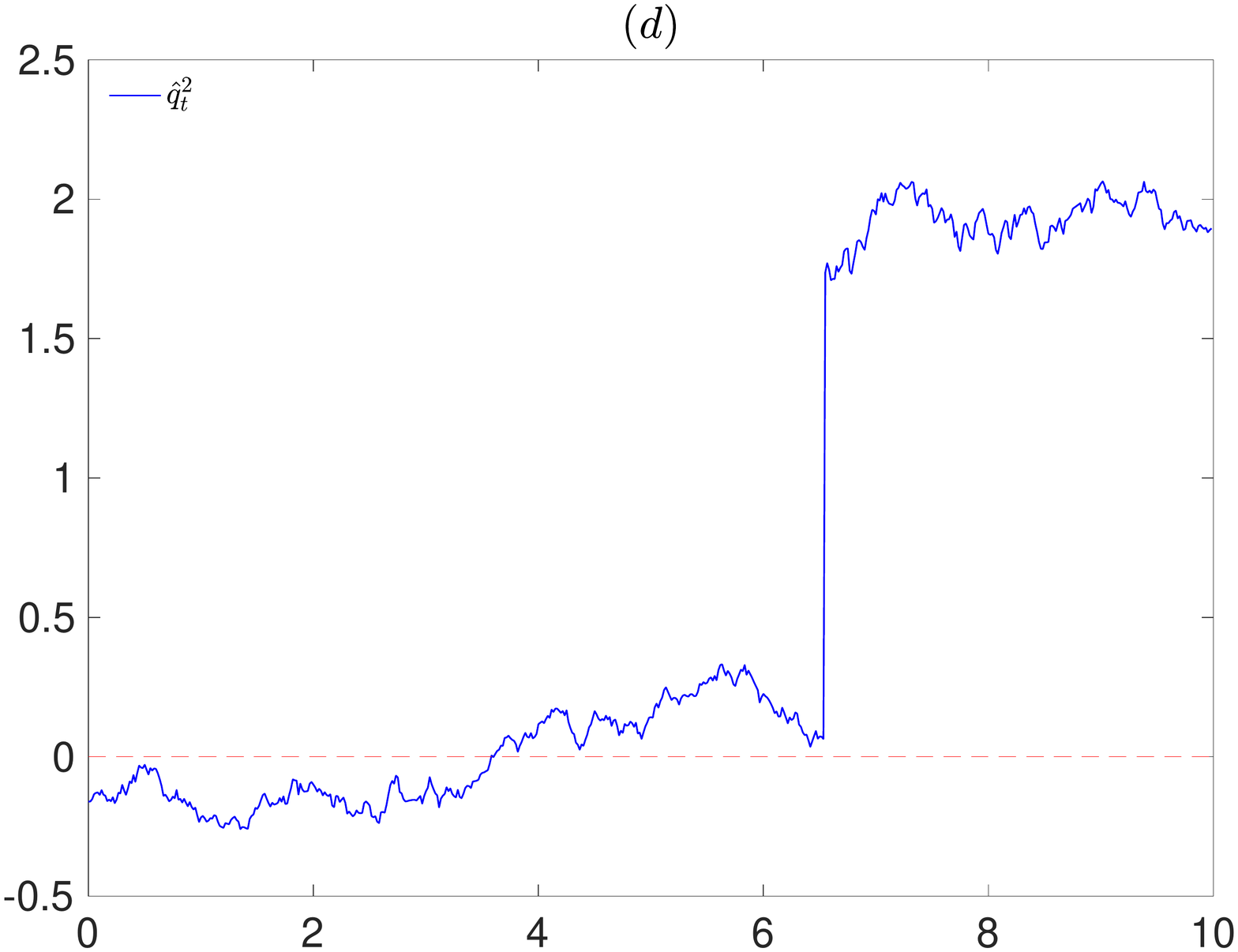}\includegraphics[width=0.3\textwidth]{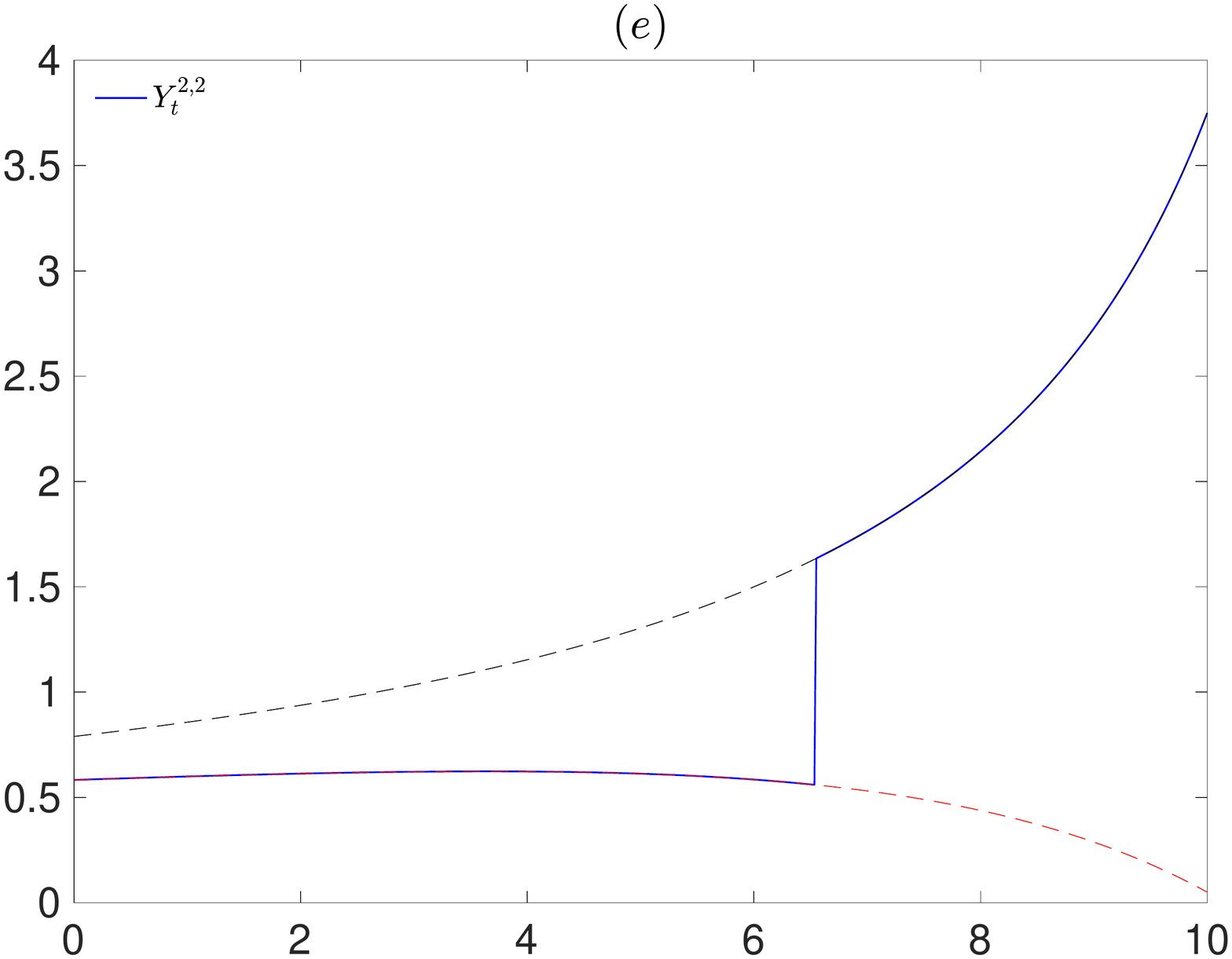}\includegraphics[width=0.3\textwidth]{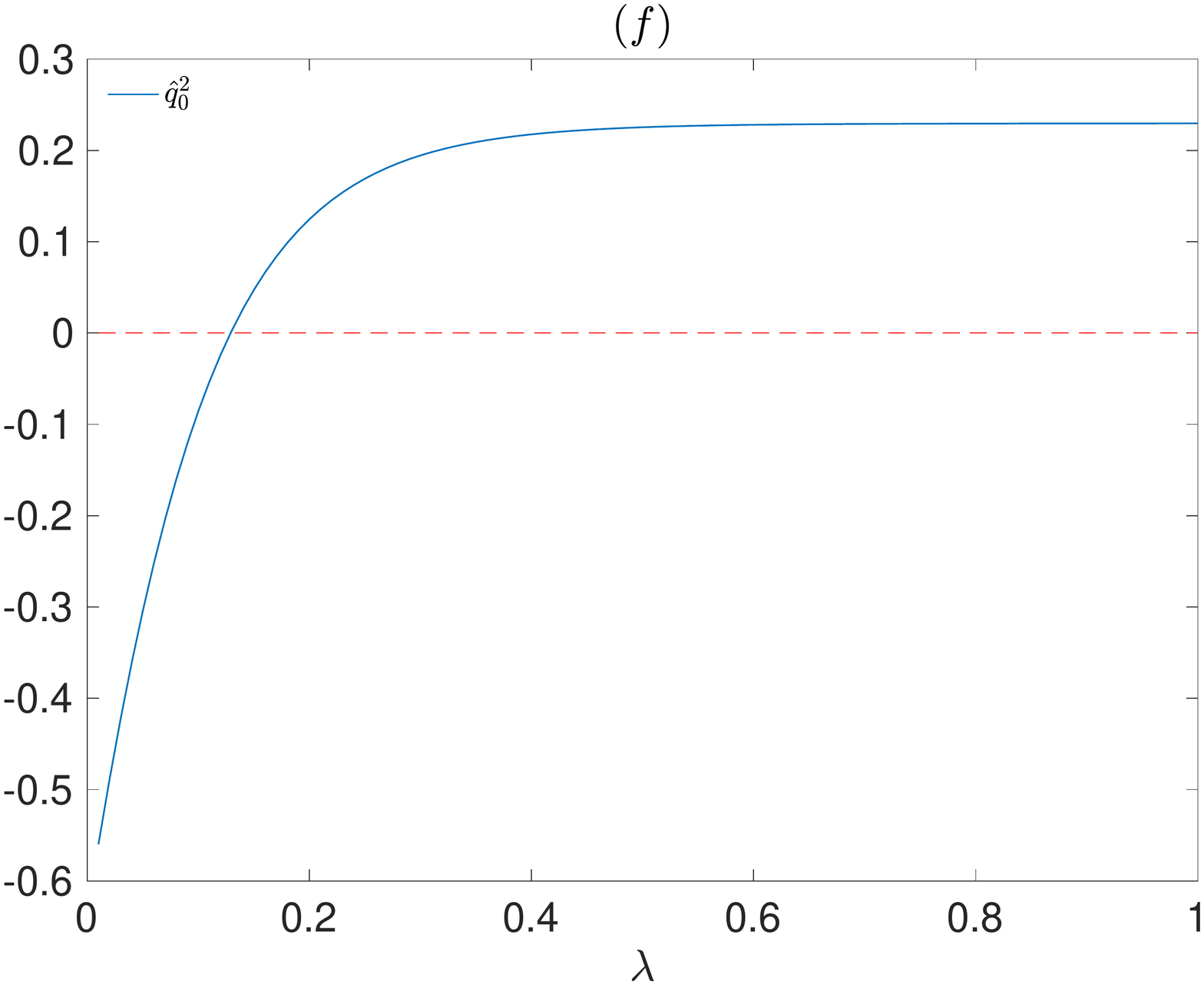}
\caption{(a) A trajectory of $\hat P_t$ with (b)  the associate trajectories of $\hat X^2_t$ and of  (c) the trading rate $\hat q^2_t$  and (e) the process $\hat Y^{2,2}_t$. (c) The evolution of $\hat P_0$ and of (f)  $\hat q^2_0$  as a function of the probability of switching from good state to bad state $\lambda = \lambda_2$.}
\label{fig:jump}
\end{figure}

\paragraph{Numerical illustration.} Figure~\ref{fig:jump} provides an illustration of the price behaviour for the case $N=2$ of Proposition~\ref{P:Jumps}. We took $\sigma^1 = \sigma^2 = 0.5$, $\sigma_0 =0$, $\rho_1 = \rho_2 = 0$, $\beta^1 = 5$ and $g = 0.1$ and $b=10$, $\eta_1 = \eta_2 = 30$ and $\lambda = 0.2$. Simulation starts in the good state for agent 2. We observe that agent 2 starts by selling power (d) because her marginal cost is lower than the marginal cost of the other agent. And when the jump occurs, she moves to a buying position. The price immediatly jumps. We also observe on figures (c) that $\hat P_0$, the price at initial time, is an increasing function of the probability of jumps. The initial trading rate of agent 2, $\hat q^2_0$, is also an increasing function of the probability of jumps. For low values, it is negative (the agent sells power) and beyond a certain threshold, she becomes a buyer.

\clearpage

\vspace{5mm}


\small
\bibliographystyle{plain}

\end{document}